\documentclass[aps,prd,reprint,amsmath,amssymb,nofootinbib,longbibliography,floatfix]{revtex4-2}
\usepackage{mathtools,physics,bm,graphicx,booktabs,microtype}
\usepackage[colorlinks=true,citecolor=blue,
linkcolor=blue,urlcolor=blue]{hyperref}
\usepackage[T1]{fontenc}
\usepackage{lmodern}
\usepackage{tikz}
\usepackage{pgfplots}
\pgfplotsset{compat=1.18}
\pgfplotsset{manuscriptaxis/.style={
 width=0.99\columnwidth,height=0.73\columnwidth,
 scale only axis=false,
 xlabel style={font=\small},ylabel style={font=\small},
 tick label style={font=\footnotesize},
 legend style={font=\footnotesize,draw=none,fill=none},
 legend cell align=left,
 every axis plot/.append style={line width=0.85pt,mark=none},
 unbounded coords=jump}}
\usetikzlibrary{decorations.pathreplacing,arrows.meta}
\usepackage{comment}
\usepackage{mathrsfs}

\graphicspath{{figures/}}
\hypersetup{pdftitle={Coherent Release Fronts in Krylov Chains and Thermal AdS2 Response},pdfauthor={Mohsen Alishahiha}}
\newcommand{\ii}{\mathrm i}
\newcommand{\e}{\mathrm e}
\newcommand{\arctanh}{\operatorname{arctanh}}
\newcommand{\cL}{\mathcal L}
\newcommand{\cP}{\mathcal P}
\newcommand{\cN}{\mathcal N}
\newcommand{\UV}{\mathrm{UV}}
\newcommand{\BF}{\mathrm{BF}}
\newcommand{\GR}{\mathrm{GR}}
\newcommand{\DG}{\Psi}
\newcommand{\hc}{\mathrm{h.c.}}

\newcommand{\cQ}{\mathsf Q}
\numberwithin{equation}{section}

\begin{document}
\raggedbottom
\clubpenalty=10000
\widowpenalty=10000
\displaywidowpenalty=10000
\title{Coherent Release Fronts in Krylov Chains and Thermal AdS$_2$ Response}
\author{Mohsen Alishahiha}
\email{alishah@ipm.ir}
\affiliation{School of Quantum Physics and Matter, Institute for Research in Fundamental Sciences (IPM), P.O. Box 19395-5531, Tehran, Iran}

\begin{abstract}
We study coherent release from a finite serial Krylov sector into an exact conformal tail with asymptotically linear Lanczos coefficients. We derive the cumulative occupation up to a cut that moves to greater Krylov depth with time, tracking the probability accumulated behind the propagating front. This quantity separates the release dynamics of the finite sector from subsequent propagation through the tail. The tail sets the logarithmic transit scale, while the finite sector shapes the resulting distribution. The early edge records the number of serial steps before emission, while the late edge reflects slow resonances and interference and can also exhibit logarithmic enhancements. Thus, one- and two-site sectors can share the same leading resonance in their continued-fraction descriptions yet produce distinct cumulative occupation patterns. We illustrate this framework in thermal AdS$_2$, where a dressed oscillator yields, in the weak-coupling limit, a two-site shallow sector coupled to the conformal tail.
\end{abstract}

\maketitle

\section{Introduction}\label{sec:intro}

Operator Krylov dynamics describes the evolution of a chosen operator as motion along a semi-infinite chain. Starting from a seed operator, repeated commutators with the Hamiltonian generate an orthonormal basis for the cyclic subspace explored by the seed. The hopping amplitudes along this chain are the Lanczos coefficients, which encode the spectral measure associated with the seed~\cite{Lanczos,viswanath1994}. Thus, the chain is determined by both the seed and the inner product.

The mean position along the chain defines operator Krylov complexity~\cite{Parker}; see Refs.~\cite{NandyReview,Rabinovici:2025otw,Jeong:2026gdc} for reviews. Spread complexity provides a related construction in state space~\cite{SpreadStates,Alishahiha:2022anw,Caputa:2024msy,Alishahiha:2022tzz}. In both cases, the resulting dynamics depends on the chosen seed and inner product, rather than on the Hamiltonian alone.

Exactly solvable conformal chains have asymptotically linear Lanczos coefficients~\cite{Parker,DymarskySmolkin,CaputaDatta}. This growth law sets a propagation scale but does not, by itself, diagnose microscopic chaos~\cite{Fullgraf,Chapman}. Subleading structure and spectral topology can provide additional dynamical information~\cite{Murugan2026}. In particular, a finite shallow sector can change frequencies, residues, interference patterns, and release delays without altering the deep tail. This raises a natural question: which features of the shallow dynamics remain visible at large Krylov depth, and how do they shape coherent emission into the tail?

Several geometric descriptions of Krylov dynamics offer complementary ways to address such questions. These include coherent-state constructions~\cite{Geometry}, bulk Hilbert-space descriptions in double-scaled SYK~\cite{LinBulk}, and relations between Krylov observables and bulk length~\cite{RabinoviciBulk,AmbrosiniLength,HellerBeyondJT}. A discrete scattering formulation maps the Krylov chain to a radial scattering problem. It relates short-depth Lanczos data to relaxation resonances and identifies logarithmic Krylov depth with a scattering coordinate~\cite{Dodelson}. Ref.~\cite{Jeong} likewise relates deep Krylov evolution to scalar propagation in a thermal AdS$_2$ throat. Together, these connections motivate a setting in which the shallow Krylov dynamics can be varied independently of the conformal propagation at large depth.

The thermal AdS$_2$ construction is governed by an $SL(2,\mathbb R)$ structure that relates the Lanczos slope to the temperature, $\alpha=\pi T$, and the representation label to the scalar mass,
\begin{equation}
m^2\ell^2=k(k-1).
\end{equation}
The choice of $k$ specifies the boundary quantization. At the Breitenlohner--Freedman (BF) bound, $k=\tfrac12$, the two scaling roots coincide, and the Lanczos coefficients are exactly linear at every depth. For general $k>0$, they remain asymptotically linear with the same leading slope. Related constructions include double-scaled SYK dictionaries~\cite{FuDSSYK} and geodesic-probe descriptions~\cite{Fatemi6d}. See also Refs.~\cite{BitaghsirFadafan:2026lek,Roychowdhury:2026igc,Nastase:2026lhz,Chatzis:2026oou,Li:2026pdh}. This conformal chain provides an exactly specified tail against which the effects of a finite shallow sector can be isolated.

To track how probability is redistributed along such a chain, we use the cumulative occupation
\begin{equation}
\cP_N(t)=\sum_{n=0}^{N}|\psi_n(t)|^2,
\label{eq:introMR}
\end{equation}
where $\psi_n(t)$ is the amplitude at site $n$ in the normalized Krylov basis. Unlike the mean position, $\cP_N(t)$ resolves how much probability lies between the seed and a specified depth. Its spectral-kernel representation and its relation to cut currents were studied in Ref.~\cite{Chowdhury}. Related cumulative Krylov quantities have been used to investigate memory~\cite{MemoryCores}, quantum Fisher information~\cite{Zhang:2025wqj,Wang:2026ysd,Alishahiha:2026qzc,adhikari2026}, symmetry-protected topological phases~\cite{MenzlerJha2026}, and higher-order temporal correlations~\cite{AlfinitoBeccaria2026}.

Our approach differs from the state-preparation protocols of Refs.~\cite{Chowdhury,MemoryCores}, which keep the Hamiltonian and reference cyclic measure fixed while changing the initial state through a polynomial preparation. That preparation induces a corresponding reweighting of the spectral measure. Here, we instead keep the seed fixed at the first site and modify a finite prefix of the Jacobi chain, leaving the deep tail unchanged. This allows us to vary the shallow Krylov dynamics directly through prescribed Jacobi coefficients and to study how its effects are carried into the conformal tail.

We implement this separation by attaching a finite ultraviolet (UV) block to an exact $SL(2,\mathbb R)$ tail through a single positive interface bond. Here, ``UV'' refers to shallow Krylov directions, rather than to a sharply defined frequency range. The model is specified directly at the Jacobi level, so a general modification of the prefix need not arise from a single fixed microscopic Hamiltonian. The tail is known exactly through its density $\rho_k$, correlation function $C_k$, propagators $f_j$, and boundary resolvent $m_k^+$. Integrating out the tail yields a boundary self-energy for the shallow block. This construction follows standard chain-mapping, projection, and memory-kernel methods~\cite{Chin,Feshbach,Mori}, as in the resonant-level~\cite{Fullgraf} and Meixner--Pollaczek~\cite{Pinna} constructions.

The cumulative occupation then provides a direct way to distinguish redistribution within the shallow block from emission into, and propagation through, the tail. The weight at the initial site tracks the return to the seed, while the total weight in the shallow block measures how much probability remains there. The occupation up to a deeper cut probes the probability that has reached the tail. These quantities must be interpreted with care: probability can continue to redistribute within the shallow block even after the seed return becomes small, and a plateau at a deep cut can simply signal delayed arrival. Moreover, probability can cross a fixed cut and later return. A fixed cut therefore need not capture the long-time propagation in a simple way. To follow the distribution as it advances to greater depth, we instead consider a cut that moves with time.

Our main result establishes the asymptotic behavior of this moving cut. For fixed positive internal bonds, interface coupling, and tail parameters, the moving-cut limit exists locally uniformly, and its derivative is given by the squared modulus of a transform of the fully coupled boundary amplitude. The tail determines the logarithmic transit scale and the propagation filter, while the resulting occupation curve retains information about the coherent emission history of the UV sector. Its early-time edge encodes the number of internal steps preceding emission. At late times, slow resonances control the power-law decay and interference, while merged poles can produce logarithmic enhancements. We also show that one- and two-site sectors can share the same leading continued-fraction pole yet exhibit distinct release profiles that cannot be related by a constant time shift.

The prescribed-tail model also admits a physically motivated realization. Building on the near-horizon comparison~\cite{Jeong} and semi-holographic constructions~\cite{Faulkner,Jensen}, we consider a boundary oscillator coupled to the operator dual to a scalar field in thermal AdS$_2$. The scalar dynamics determines the retarded response of this operator, whose thermal spectral measure fixes the conformal tail. Although this construction applies for general $k>0$, we focus on the BF bound, $k=1/2$, where the ingoing bulk solution and the renormalized source variation yield a particularly simple response.

Integrating out the boundary operator gives the dressed oscillator measure, which determines the oscillator's effective Krylov chain. In the weak-coupling limit, its moments approach those of a two-site shallow block followed by the conformal tail at fixed depth. The interface normalization and first-bond correction also reproduce the leading complex oscillator pole through order $\lambda^2$.

This realization assumes a fixed scalar background and a bath that remains conformal at all frequencies. We do not include gravitational backreaction or Schwarzian fluctuations.

The paper is organized as follows. Section~\ref{sec:thermalK} introduces the observable and the conformal reference chain. Section~\ref{sec:UV} develops the coherent kernel, establishes its moving-cut limit, and derives the edge asymptotics. Section~\ref{sec:examples} studies minimal UV sectors and compares systems with matched leading poles. Section~\ref{sec:gravity} derives the oscillator parameters from the BF response and establishes the leading-pole matching. The final section presents our conclusions. Technical derivations of the spectral kernel and bulk response are given in Appendices~\ref{app:front} and~\ref{app:bulk}.


\section{Thermal Krylov dynamics and the conformal reference}\label{sec:thermalK}

This section sets the operator-space conventions and Lanczos recursion. We then define the occupation probability and its continuity equation, and give the conformal chain in closed form.

\subsection{Thermal operator space and Lanczos evolution}

We work in the thermal operator space, equipped with the symmetric
inner product
\begin{equation}
 (A|B)_\beta=Z_\beta^{-1}\Tr\!\left(\e^{-\beta H/2}A^\dagger\e^{-\beta H/2}B\right),
 \label{eq:product}
\end{equation}
for a Hamiltonian $H$ and operators $A,B$, with
$Z_\beta=\Tr\e^{-\beta H}$. This symmetric splitting is the convention used throughout this paper. It makes the Liouvillian
$\cL=[H,\cdot]$ self-adjoint and gives the seed a positive norm. For a
Hermitian seed $O$, write $\cN_\beta=(O|O)_\beta>0$ and
$O_0=O/\sqrt{\cN_\beta}$. Hermiticity of the seed and the symmetric
splitting make the associated measure even, so Lanczos
orthogonalization has no diagonal coefficient~\cite{viswanath1994}
\begin{equation}
 \cL|O_n)=b_{n+1}|O_{n+1})+b_n|O_{n-1}),\qquad b_0=0.\label{eq:lanczos}
\end{equation}
We identify this cyclic representation with a real Jacobi operator $K$
acting on the Krylov basis $\{e_n\}_{n\ge0}$, where $e_0=O_0$ is the
normalized seed and $e_n$ for $n\ge1$ are the subsequent Lanczos
vectors. We use the scattering convention
\begin{align}
 |O(t))&=\e^{- i  Kt}|e_0\rangle,\qquad \psi_n=(- i )^n\varphi_n,\nonumber\\
 \dot\varphi_n&=b_n\varphi_{n-1}-b_{n+1}\varphi_{n+1},\qquad \varphi_n(0)=\delta_{n0}.\label{eq:chain}
\end{align}
The amplitudes $\varphi_n$ are real for this zero-diagonal chain and
seed.

The unnormalized Heisenberg-evolved seed, setting
$\hbar=1$, is
$O(t)=e^{ i  Ht}Oe^{- i  Ht}$. Its even symmetric-split correlation
equals the return amplitude in the scattering convention,
\begin{align}
 C_W(t)&=\frac{\Tr(\e^{-\beta H/2}O(t)\e^{-\beta H/2}O)}{Z_\beta\cN_\beta}
 =\int\e^{- i \omega t}d \mu_W(\omega),\label{eq:CW}\\
 d \mu_W(\omega)&=\frac1{Z_\beta\cN_\beta}\sum_{a,b}
 \e^{-\beta(E_a+E_b)/2}|O_{ab}|^2\nonumber\\[-2pt]
 &\hspace{14mm}
 \times\delta(\omega-E_b+E_a)d \omega,\label{eq:Lehmann}
\end{align}
where $O_{ab}=\langle E_a|O|E_b\rangle$ is the matrix element in the
energy eigenbasis, $H|E_a\rangle=E_a|E_a\rangle$. In the absolutely
continuous case relevant here, the measure is written as
$d \mu_W(\omega)=\rho_W(\omega)d \omega$, with $\rho_W$ the
normalized symmetric-split Wightman density.

The orthonormal polynomials
associated with $\mu_W$ satisfy \cite{Szego1975,DLMF18,MuckYang}
\begin{align}
 \omega P_n&=b_{n+1}P_{n+1}+b_nP_{n-1},\quad P_{-1}=0,\ P_0=1,\label{eq:polyrec}\\
 \psi_n(t)&=\int P_n(\omega)\e^{- i \omega t}d \mu_W(\omega)
 =P_n( i \partial_t)C_W(t).\label{eq:spectralamplitude}
\end{align}
Thus the full chain is fixed by the measure $\mu_W$, and the amplitudes
$\psi_n$ are its orthogonal-polynomial moments.

The Lanczos coefficients are determined by the moments of the normalized seed measure. For an even measure, define
$
\mu_j=\int \omega^j\,d\mu_W(\omega),
$ so that $\mu_0=1$ and $\mu_{2p+1}=0$. Equivalently,
\begin{equation}
  \mu_{2p}=(-1)^p C_W^{(2p)}(0)
=(O_0|\cL^{2p}O_0)_\beta.  
\end{equation}
These moments are therefore fixed by the short-time seed correlation, or equivalently by successive commutators with the Hamiltonian.

More generally, introduce the Hankel determinants
\begin{equation}
   \mathsf H_n=\det[\mu_{i+j}]_{i,j=0}^{n-1},
\qquad
\mathsf H_0=1. 
\end{equation}
Orthogonalizing the monomials gives the squared norm $\mathsf H_{n+1}/\mathsf H_n$ of the monic polynomial of degree $n$, yielding
\begin{equation}
b_n^2=
\frac{\mathsf H_{n+1}\mathsf H_{n-1}}{\mathsf H_n^2},
\qquad n\ge1,
\label{eq:hankelbonds}
\end{equation}
provided the relevant Hankel determinants are positive. For example,
\begin{equation}
b_1^2=\mu_2,\qquad
b_2^2=\frac{\mu_4-\mu_2^2}{\mu_2},\qquad
 b_3^2=
\frac{\mu_2\mu_6-\mu_4^2}
{\mu_2(\mu_4-\mu_2^2)}.
\label{eq:momentbonds}
   \end{equation}
The last expression assumes $b_1,b_2\neq0$. If the first vanishing Hankel determinant occurs at $\mathsf H_{n+1}$ while $\mathsf H_0,\ldots,\mathsf H_n$ remain positive, the Krylov recursion terminates at depth $n$.

\subsection{Cumulative weight, continuity, and early-time propagation}\label{sec:spectral}

We now define the cumulative occupation probability. Let $\Pi_N$ denote the projector onto the first $N+1$ Krylov directions,
\begin{equation}
\Pi_N=\sum_{n=0}^{N}|e_n\rangle\langle e_n|,
\label{eq:projector}
\end{equation}
so that the cumulative occupation probability is~\cite{Chowdhury,MemoryCores}
\begin{equation}
\cP_N(t)=|\Pi_N|\psi(t)\rangle|^2
=\sum_{n=0}^{N}\varphi_n^2(t),
\label{eq:MR}
\end{equation}
which measures the total weight contained in the Krylov directions from the seed through depth $N$. Unlike the mean Krylov depth, which tracks the average position of the distribution, $\cP_N$ probes the weight within a specified region of the chain.

A large $\cP_N(t)$ at early times can have different origins, such as slow escape through a weak interface, internal oscillations within a finite sector, or delayed propagation to a distant cut. A single decay rate cannot distinguish these mechanisms, whereas comparing the seed, shallow-sector, and deeper-cut occupations can.

The projection form also gives
\begin{equation}
  0\le \cP_N(t)\le1,\qquad
\cP_{N+1}(t)\ge\cP_N(t),  
\end{equation}
at every fixed time, with
$\cP_N(t)\longrightarrow1$ for $N\to\infty$.

Differentiating Eq.~\eqref{eq:MR} and using the chain equation~\eqref{eq:chain}, the internal bond contributions cancel pairwise, giving
\begin{equation}
\dot\cP_N(t)
=-2b_{N+1}\varphi_N(t)\varphi_{N+1}(t).
\label{eq:continuity}
\end{equation}
This is the discrete continuity equation for probability transport in Krylov space. Defining the current through the bond between sites $N$ and $N+1$ by
\begin{equation}
J_{N+1}=2b_{N+1}\varphi_N\varphi_{N+1},
\label{eq:current}
\end{equation}
we have $\dot\cP_N=-J_{N+1}$, with $J_{N+1}>0$ denoting outward flow.

The current can change sign, allowing weight to return from deeper Krylov directions. Consequently, $\cP_N(t)$ need not decrease monotonically. Since $J_{N+1}$ depends on the product of neighboring amplitudes, it captures coherent transport rather than an incoherent escape rate.

Using Eq.~\eqref{eq:spectralamplitude}, the cumulative probability reads
\begin{equation}
\cP_N(t)=\iint e^{- i (\omega-\omega')t}
\mathsf K_N(\omega,\omega')\,
d\mu_W(\omega)d\mu_W(\omega'),
\label{eq:MRspectral}
\end{equation}
with
\begin{equation}
\mathsf K_N(\omega,\omega')
=\sum_{n=0}^{N}P_n(\omega)P_n(\omega').
\label{eq:kernel}
\end{equation}
The kernel is determined entirely by the full measure and does not select an independent frequency window.

Because the chain~\eqref{eq:chain} has vanishing diagonal entries, $\cP_N(t)$ is even in $t$. For fixed $N$,
\begin{equation}
1-\cP_N(t)
=\frac{(b_1\cdots b_{N+1})^2}{[(N+1)!]^2}
t^{2N+2}+O(t^{2N+4}),
\label{eq:shorttime}
\end{equation}
as $t\to0$. Thus, a distant cut remains initially flat even without a slow UV mode. This fixed-depth expansion, however, does not determine the plateau duration at large $N$: its validity need not extend to a depth-dependent crossing time. The exact solution below provides the relevant time scale.

Finally, since $\varphi_N^2=\cP_N-\cP_{N-1}$ with $\cP_{-1}=0$, the family ${\cP_N}$ determines the full depth distribution. The mean Krylov depth then follows from the tail-sum identity~\cite{Chowdhury},
\begin{equation}
\mathcal C_K(t)
=\sum_{n\ge0}n\varphi_n^2(t)
=\sum_{N\ge0}[1-\cP_N(t)],
\label{eq:complexity}
\end{equation}
whenever the sums converge.

\subsection{The exact conformal propagation reference}\label{sec:reference}

We now introduce a solvable reference in which the spectral measure and
chain amplitudes are known exactly. The $SL(2,\mathbb R)$ discrete-series
chain provides the required propagation benchmark and makes the
plateau, transit, and falloff regimes explicit.

Its hopping amplitudes are~\cite{Parker,DymarskySmolkin,CaputaDatta}
\begin{equation}
b_j=\alpha\sqrt{j(j+2k-1)},\qquad j\ge1,\quad \alpha,k>0.
\label{eq:tailbonds}
\end{equation}
The chain is generated by $K=\alpha(L_++L_-)$ in a positive
discrete-series representation of $\mathfrak{sl}(2,\mathbb R)$ with
Casimir $k(k-1)$. The parameter $\alpha$ sets the asymptotic growth
scale, while $k$ labels the representation and determines the full
sequence of bonds. Here $L_\pm$ are the
raising and lowering generators of 
$\mathfrak{sl}(2,\mathbb R)$.

The corresponding seed correlation, spectral density, and
phase-stripped propagator are~\cite{DymarskySmolkin,CaputaDatta,Geometry,Gamayun}
\begin{align}
C_k(t)&=\sech^{2k}(\alpha t),\label{eq:tailC}\\
\rho_k(\omega)&=\frac{2^{2k-2}}{\pi\alpha\Gamma(2k)}
\left|\Gamma\left(k+\frac{ i \omega}{2\alpha}\right)\right|^2,
\label{eq:tailrho}\\
f_j(t)&=\sqrt{\frac{(2k)_j}{j!}}\,
\tanh^j(\alpha t)\sech^{2k}(\alpha t).
\label{eq:tailf}
\end{align}
Here $C_k(t)=\langle0|\e^{- i  Kt}|0\rangle$ is the seed correlation,
$\rho_k(\omega)$ is the seed spectral density, and $f_j(t)$ is defined by
\begin{equation}
\langle j|\e^{- i  Kt}|0\rangle=(- i )^j f_j(t).
\label{eq:tailprop}
\end{equation}
The spectral density is smooth and strictly positive on the real axis,
and the measure is purely absolutely continuous, with no bound-state
contribution.

Introducing $x=\tanh^2(\alpha t)$, the site occupation is
\begin{equation}
|f_j(t)|^2=
\frac{(2k)_j}{j!}(1-x)^{2k}x^j,
\label{eq:pj}
\end{equation}
a negative-binomial distribution in $j$. Hence
\begin{equation}
\cP_N(t)=\sum_{j=0}^N|f_j(t)|^2
=I_{1-x}(2k,N+1),
\label{eq:baregeneral}
\end{equation}
where $I_z(a,b)$ denotes the regularized incomplete beta function.

For $k=\tfrac12$, the bonds become exactly linear,
$b_j=\alpha j$, and the cumulative occupation reduces to
\begin{equation}
\cP_N(t)=1-\tanh^{2(N+1)}(\alpha t).
\label{eq:bareBF}
\end{equation}
This is the BF case used below. For fixed $N$ and $\alpha t\ll1$,
\begin{equation}
1-\cP_N(t)
=(\alpha t)^{2N+2}
+O((\alpha t)^{2N+4}),
\label{eq:BFshorttime}
\end{equation}
which realizes the general short-time behavior in
Eq.~\eqref{eq:shorttime}. The resulting initial flatness is therefore a
fixed-depth effect.

The duration of this plateau is instead set by propagation to the cut.
For a fixed $0<\delta<1$, define $t_\delta(N)$ by
$\cP_N(t_\delta)=\delta$. Then
\begin{equation}
\begin{aligned}
t_\delta(N)
&=\frac1\alpha\arctanh\left[(1-\delta)^{1/[2(N+1)]}
\right]\\
&=\frac1{2\alpha}\log\frac{4(N+1)}{L_\delta}
+\frac{L_\delta^2}{96\alpha(N+1)^2}\\
&\;\;\;\;\;+O\left(\frac1{\alpha(N+1)^4}\right),
\end{aligned}
\label{eq:crossingbare}
\end{equation}
where $L_\delta=-\log(1-\delta)$. Thus, for any fixed plateau threshold
$0<\epsilon\ll1$, defined by $1-\cP_N=\epsilon$, the plateau duration
grows as $\log N/(2\alpha)$. This logarithmic delay reflects propagation
through the conformal tail rather than a slow localized mode.

The exact propagator and everywhere-positive continuous measure therefore
provide a clean reference with no localized spectral component. This
property belongs to the specified conformal model and is not a general
assumption about Krylov chains.

\section{A finite ultraviolet sector coupled to an exact conformal tail}\label{sec:UV}

The exact $SL(2,\mathbb R)$ chain of Sec.~\ref{sec:reference} fixes
propagation along the conformal tail but not the dynamics of the initial
Krylov directions. We therefore couple a finite UV sector to the exact
tail and derive a closed expression for the cumulative occupation
probability at arbitrary depth. Throughout, the seed remains at site
zero, while the finite prefix is varied with the entire tail held fixed.

\subsection{The model}\label{subsec:model}

We now specify the Jacobi model used throughout this section. Its finite
prefix encodes the tunable UV dynamics, while the remaining coefficients
are fixed to those of the conformal tail in Sec.~\ref{sec:reference}.
The construction is defined directly at the Jacobi level, so a chosen
UV prefix need not correspond to the Krylov chain of a single fixed
microscopic Hamiltonian.

The UV sector contains $d\ge1$ sites, labeled
$0,\ldots,d-1$, with positive internal bonds
$b_1^{\UV},\ldots,b_{d-1}^{\UV}$,
\begin{equation}
K_\UV=\sum_{j=1}^{d-1}b_j^{\UV}
\bigl(|e_{j-1}\rangle\langle e_j|+\hc\bigr),
\qquad b_j^{\UV}>0 ,
\label{eq:UVblock}
\end{equation}
where $|e_j\rangle$ denotes the $j$-th UV Krylov basis vector. We attach
the exact conformal chain of Sec.~\ref{sec:reference}, which we refer to
as the tail, at global site $d$ through a positive interface bond $g$:
\begin{equation}
K=\begin{pmatrix}
K_\UV & g\,|e_b\rangle\langle 0_T|\\[2pt]
g\,|0_T\rangle\langle e_b| & K_T
\end{pmatrix},
\qquad g>0 ,
\label{eq:blockK}
\end{equation}
where $|e_b\rangle=|e_{d-1}\rangle$ is the boundary UV vector and
$|0_T\rangle=|e_d\rangle$ is the tail origin. The tail bonds
$b_j^{\mathrm T}$ are those of Eq.~\eqref{eq:tailbonds}. In global
labels, the full sequence of bonds is
\begin{equation}
b_j=
\begin{cases}
b_j^{\UV}, & 1\le j<d,\\[2pt]
g, & j=d,\\[2pt]
b_{j-d}^{\mathrm T}, & j>d .
\end{cases}
\label{eq:fullbonds}
\end{equation}
Thus, $g$ is the $d$-th Lanczos coefficient of the full chain, while
$b_j^{\UV}$ are the preceding $d-1$ coefficients. We treat $g$
separately because it is the sole bond connecting the two sectors.

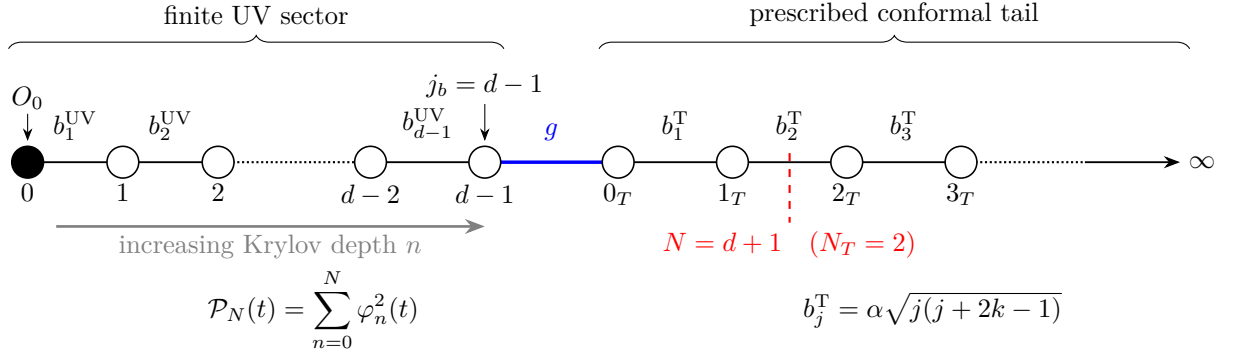
\begin{figure*}[t]
\centering
\begin{tikzpicture}[x=1.26cm,y=1cm,>=Stealth,
every node/.style={font=\normalsize},site/.style={circle,draw,fill=white,
minimum size=4.2mm,inner sep=0pt,line width=.6pt}]
\draw[line width=.7pt] (0,0)--(2,0);
\draw[densely dotted,line width=.8pt] (2,0)--(3.6,0);
\draw[line width=.7pt] (3.6,0)--(4.8,0);
\draw[blue,line width=1.3pt] (4.8,0)--(6.2,0);
\draw[line width=.7pt] (6.2,0)--(9.8,0);
\draw[densely dotted,line width=.8pt] (9.8,0)--(11.1,0);
\draw[->,line width=.7pt] (11.1,0)--(12.1,0);
\node[site,fill=black] at (0,0) {};
\foreach \x in {1,2,3.6,4.8,6.2,7.4,8.6,9.8} \node[site] at (\x,0) {};
\node[below=5pt] at (0,0) {$0$};
\node[below=5pt] at (1,0) {$1$};
\node[below=5pt] at (2,0) {$2$};
\node[below=5pt] at (3.6,0) {$d-2$};
\node[below=5pt] at (4.8,0) {$d-1$};
\node[below=5pt] at (6.2,0) {$0_T$};
\node[below=5pt] at (7.4,0) {$1_T$};
\node[below=5pt] at (8.6,0) {$2_T$};
\node[below=5pt] at (9.8,0) {$3_T$};
\node[right] at (12.1,0) {$\infty$};
\node[above=5pt] at (.5,0) {$b_1^{\UV}$};
\node[above=5pt] at (1.5,0) {$b_2^{\UV}$};
\node[above=5pt] at (4.2,0) {$b_{d-1}^{\UV}$};
\node[above=5pt,blue] at (5.5,0) {$g$};
\node[above=5pt] at (6.8,0) {$b_1^{\mathrm T}$};
\node[above=5pt] at (8,0) {$b_2^{\mathrm T}$};
\node[above=5pt] at (9.2,0) {$b_3^{\mathrm T}$};
\node at (0,.85) {$O_0$}; \draw[->] (0,.64)--(0,.28);
\node at (4.8,1.02) {$j_b=d-1$}; \draw[->] (4.8,.77)--(4.8,.28);
\draw[decorate,decoration={brace,amplitude=5pt}] (-.2,1.5)--(5,1.5);
\node at (2.4,1.95) {finite UV sector};
\draw[decorate,decoration={brace,amplitude=5pt}] (6,1.5)--(12.2,1.5);
\node at (9.1,1.95) {prescribed conformal tail};
\draw[red,dashed,line width=.8pt] (8,.12)--(8,-.8);
\node[red,below] at (8,-.8) {$N=d+1\quad(N_T=2)$};
\draw[->,gray,line width=1pt] (.3,-.85)--(4.8,-.85);
\node[gray,below] at (2.55,-.85) {increasing Krylov depth $n$};
\node at (3,-1.95) {$\cP_N(t)=\displaystyle\sum_{n=0}^N\varphi_n^2(t)$};
\node at (9.5,-1.95) {$b_j^{\mathrm T}=\alpha\sqrt{j(j+2k-1)}$};
\end{tikzpicture}
\caption{A finite UV sector coupled through $g$ (blue) to a prescribed
conformal tail. The red observation boundary is drawn after $1_T$:
the illustrated cut includes $N_T=2$ tail sites, so its last global
index is $N=d+1$. It is a projection boundary, not an absorbing wall.
The chain remains discrete; a smooth-envelope description in logarithmic
depth applies only at large index.}
\label{fig:UV-Tail}
\end{figure*}

The seed is $|O_0)=|e_0)$ in the operator-space notation of
Sec.~\ref{sec:thermalK}, and the tail is initially empty. If we include
$N_T\ge0$ tail sites, the observation cut lies at
$N=d+N_T-1$. Hence, $\cP_{d-1}$ gives the total weight in the UV sector,
while $\cP_{d+N_T-1}$ includes the first $N_T$ tail sites, as illustrated
in Fig.~\ref{fig:UV-Tail}.

This choice of tail provides several useful features. Its spectral
measure, correlation function, and propagator are known in closed form,
so the tail can be integrated out exactly. Its bonds are asymptotically
linear, as in the near-horizon construction of Ref.~\cite{Jeong}, while
its seed measure is continuous and strictly positive on $\mathbb R$ and
therefore contains no bound component that could mimic localization.
The UV bonds control the shallow frequencies and low moments, whereas
the interface $g$ controls the exchange between the two sectors. We can
therefore vary $d$, $b_j^{\UV}$, and $g$ while keeping the conformal tail
unchanged, providing a controlled setting in which to isolate the role
of the finite UV dynamics.


\subsection{Eliminating the tail in frequency space}\label{subsec:integrate}

The conformal tail is infinite, but it couples to the UV sector through a
single boundary site. Its effect on the UV dynamics is therefore encoded
in the boundary resolvent
\begin{equation}
m_k^+(z)=\langle 0_T|(z-K_T)^{-1}|0_T\rangle,
\qquad \Im z>0 .
\label{eq:tailm}
\end{equation}
For the exact $SL(2,\mathbb R)$ chain of Sec.~\ref{sec:reference}, this
resolvent can be written in closed form in terms of the incomplete beta
function. Using Eq.~\eqref{eq:tailC},
\begin{align}
m_k^+(z)&=-\frac{ i  2^{2k-1}}{\alpha}
B_{1/2}\!\left(k-\frac{ i  z}{2\alpha},
k+\frac{ i  z}{2\alpha}\right),
\label{eq:mkgeneral}\\
m_{1/2}^+(z)&=-\frac{ i }{2\alpha}\left[
\DG\!\left(\frac{3- i  z/\alpha}{4}\right)
-\DG\!\left(\frac{1- i  z/\alpha}{4}\right)\right],
\label{eq:mkBF}
\end{align}
where $B$ is the incomplete beta function and $\DG$ denotes the digamma
function. The superscript $+$ indicates continuation from the upper
half-plane. On the real axis,
\begin{equation}
\Im m_k^+(\omega)=-\pi\rho_k(\omega),
\label{eq:mimag}
\end{equation}
so the imaginary part of the boundary resolvent directly gives the
spectral density of the tail.

The Schur complement of the tail block in $zI-K$ then replaces the
infinite tail by a rank-one self-energy at the UV boundary:
\begin{equation}
G_{\UV}(z)=
\bigl[zI_d-K_\UV-g^2m_k^+(z)\,e_b e_b^\dagger\bigr]^{-1}.
\label{eq:GUV}
\end{equation}
Thus, all effects of the tail enter the finite UV problem through the
scalar self-energy $g^2m_k^+(z)$. Its real part shifts the UV response,
while its imaginary part accounts for spectral weight flowing into the
tail. In the decoupled limit $g=0$, the self-energy vanishes and
$G_{\UV}$ reduces to the bare UV resolvent.

The seed spectral density is obtained from the $(0,0)$ component of this
dressed resolvent,
\begin{equation}
\rho_\UV^{(d)}(\omega)
=-\frac1\pi\Im[G_{\UV}]_{00}(\omega+ i 0).
\label{eq:spectralconvention}
\end{equation}
On the real axis, the dressed UV resolvent obeys
\begin{equation}
-\Im[G_{\UV}]_{00}
=\pi g^2\rho_k(\omega)\,
\bigl|[G_{\UV}]_{0b}\bigr|^2 .
\label{eq:optical}
\end{equation}
This follows from the identity
$\Im G=-G\,\Im(G^{-1})\,G^\dagger$ applied to
$G_{\UV}^{-1}$ together with Eq.~\eqref{eq:mimag}. The identity
determines the absorptive part, but the full complex resolvent is still
needed to capture the coherent dynamics.

We next express the dressed resolvent in terms of a simple determinant
recursion. Let $\Delta_j(z)$ denote the determinant of the leading
$j\times j$ principal submatrix of $zI_d-K_\UV$. Tridiagonality gives
\begin{align}
& \det\!\left[zI_d-K_{\UV}
 -g^2m_k^+(z)e_be_b^\dagger\right]
 =
 \Delta_d(z)\nonumber\\
&\qquad\times\left[
 1-g^2m_k^+(z)
 e_b^\dagger(zI_d-K_{\UV})^{-1}e_b
 \right].
 \label{eq:detlemma}
\end{align}
The relevant boundary cofactor is $\Delta_{d-1}$, giving
\begin{equation}
e_b^\dagger(zI_d-K_{\UV})^{-1}e_b
=\frac{\Delta_{d-1}(z)}{\Delta_d(z)} .
\label{eq:barematrixelement}
\end{equation}
The dressed determinant therefore takes the compact form
\begin{equation}
\mathcal D_d(z)=\Delta_d(z)-g^2m_k^+(z)\Delta_{d-1}(z).
\label{eq:denominator}
\end{equation}
Although the intermediate expression assumes $\Delta_d(z)\neq0$, this final
determinant identity extends through the zeros of $\Delta_d$ by analytic
continuation.

To obtain the seed-to-boundary matrix element, define
\begin{equation}
U=\prod_{j=1}^{d-1}b_j^{\UV},
\end{equation}
with $U=1$ for $d=1$. Because the UV block is a chain, the cofactor
connecting sites $0$ and $d-1$ contains a unique path through all
intervening bonds, whose magnitude is $U$. Cramer's rule then gives
\begin{equation}
[G_{\UV}]_{0b}(z)=[G_{\UV}]_{b0}(z)
=\frac{U}{\mathcal D_d(z)} .
\label{eq:G0b}
\end{equation}
Combining this result with the optical identity yields the seed spectral
density
\begin{equation}
\rho_{\UV}^{(d)}(\omega)
=\frac{g^2U^2\rho_k(\omega)}
{|\mathcal D_d(\omega)|^2},
\label{eq:UVdensity}
\end{equation}
where the empty product is understood as one for $d=1$. The structure is
transparent: $g^2\rho_k(\omega)$ gives the spectral weight supplied by
the tail through the interface, $U^2$ is a frequency-independent
prefactor fixed by the serial UV bonds, and the dressed determinant
$\mathcal D_d(\omega)$ contains the nontrivial frequency dependence of
the finite sector.

The zeros of $\mathcal D_d(z)$ determine the dressed resonances of the
coupled system. Their continuation into the complex plane controls the
late-time boundary history and, consequently, the late-edge behavior
discussed in Sec.~\ref{sec:edges}. These resonances are poles of the
analytically continued resolvent rather than eigenvalues of the
self-adjoint chain. In the decoupling limit $g\to0$,
$\mathcal D_d\to \Delta_d$, so the dressed resonances reduce to the bare UV
frequencies; for nonzero $g$, they acquire the corresponding shifts and
broadening induced by the tail.


\subsection{Memory equation and coherent release in time}\label{subsec:memory}

The previous subsection eliminated the tail in frequency space. We now
eliminate it in the time domain, reducing the coupled problem to a
finite-dimensional equation for the UV sector. The tail then appears
through an explicit memory kernel, while the cumulative occupation
probability becomes a double integral over the boundary emission history.
This representation makes the coherent release mechanism transparent.

Consider the normalized scattering-convention state
$|O(t))=\e^{- i  Kt}|e_0\rangle$, decomposed into the two sectors,
\begin{equation}
\begin{aligned}
|O(t))&=\sum_{j=0}^{d-1}\xi_j^{\UV}(t)|e_j\rangle
+\sum_{j=0}^{\infty}\xi_j^{\mathrm T}(t)|j_T\rangle,\\
\xi_j^{\UV}(0)&=\delta_{j0},\qquad
\xi_j^{\mathrm T}(0)=0,
\end{aligned}
\label{eq:decomposition}
\end{equation}
where $|j_T\rangle=|e_{d+j}\rangle$ is the $j$-th tail basis vector. We
write $\xi_b\equiv\xi_{d-1}^{\UV}$ for the UV boundary amplitude. Thus
$\xi_j^{\UV}=\psi_j$ and $\xi_j^{\mathrm T}=\psi_{d+j}$.

Projecting $ i \partial_t|O(t))=K|O(t))$ onto the two sectors gives
\begin{align}
 i \dot\xi_j^{\UV}
&=\sum_{j'=0}^{d-1}(K_\UV)_{jj'}\xi_{j'}^{\UV}
+g\,\delta_{j,d-1}\,\xi_0^{\mathrm T},
\label{eq:UVeq}\\
 i \dot\xi_j^{\mathrm T}
&=\sum_{j'=0}^{\infty}(K_T)_{jj'}\xi_{j'}^{\mathrm T}
+g\,\delta_{j,0}\,\xi_b .
\label{eq:driventail}
\end{align}
The tail is therefore driven only through its boundary by $\xi_b$. With
$\xi^{\mathrm T}(0)=0$, Duhamel's formula gives
\begin{equation}
\xi_j^{\mathrm T}(t)
=- i  g(- i )^j
\int_0^t f_j(t-s)\,\xi_b(s)\,ds ,
\label{eq:emission}
\end{equation}
where $f_j$ is the real phase-stripped propagator defined by
$[e^{- i  K_T\tau}]_{j0}=(- i )^j f_j(\tau)$. Substituting this solution
into the UV equation yields the exact
finite-dimensional memory equation
\begin{equation}
 i \dot\xi_j^{\UV}
=\sum_{j'=0}^{d-1}(K_\UV)_{jj'}\xi_{j'}^{\UV}
- i  g^2\delta_{j,d-1}
\int_0^t C_k(t-s)\,\xi_b(s)\,ds ,
\label{eq:memory}
\end{equation}
with
\begin{equation}
C_k(\tau)=\langle0_T|\e^{- i  K_T\tau}|0_T\rangle
=\int_{\mathbb R}\e^{- i \omega\tau}\rho_k(\omega)\,d\omega .
\label{eq:Ck_kernel}
\end{equation}
Thus the boundary dynamics depends on its entire past, weighted by the
tail return amplitude. In frequency space, the same effect is encoded by
the self-energy $g^2m_k^+$. If the boundary amplitude varies slowly on
the tail memory scale and is concentrated near zero frequency, the local
approximation gives the leading amplitude-damping coefficient
$\pi g^2\rho_k(0)$. An oscillatory UV mode instead samples the response
at its characteristic frequency.

Importantly, $\xi_b(s)$ in Eq.~\eqref{eq:emission} is the boundary
amplitude of the fully coupled system, so it already includes the
backaction of previously emitted amplitude. The integral therefore sums
amplitudes, not probabilities: emissions at different times interfere.
This coherent history is the origin of the backflow and non-monotonic
finite-cut occupation discussed below.

Norm conservation gives
\begin{equation}
\sum_{j=0}^{d-1}|\xi_j^{\UV}(t)|^2
+\sum_{j=0}^{\infty}|\xi_j^{\mathrm T}(t)|^2=1 .
\label{eq:normalization}
\end{equation}
For a cut after $N_T$ tail sites, $N=d+N_T-1$, so
\begin{equation}
\cP_{d+N_T-1}(t)
=1-\sum_{j=N_T}^{\infty}|\xi_j^{\mathrm T}(t)|^2 .
\label{eq:retention_complement}
\end{equation}
Using Eq.~\eqref{eq:emission} and exchanging the sum with the time
integrals gives
\begin{equation}
\sum_{j=N_T}^{\infty}|\xi_j^{\mathrm T}(t)|^2
=g^2\int_0^t ds\int_0^t ds'\,
\xi_b(s)\xi_b(s')^*\,K_{N_T}(t;s,s'),
\label{eq:omitted_kernel}
\end{equation}
where
\begin{equation}
K_{N_T}(t;s,s')
=\sum_{j=N_T}^{\infty}f_j(t-s)f_j(t-s').
\label{eq:KN_def}
\end{equation}

For the exact $SL(2,\mathbb R)$ tail, this kernel can be evaluated
explicitly. Define
\begin{equation}
Q_t(s,s')
=\tanh[\alpha(t-s)]\tanh[\alpha(t-s')]
.
\label{eq:Qt}
\end{equation}
Using the product form of the tail propagators together with the
negative-binomial series gives
\begin{equation}
\sum_{j=N_T}^{\infty}f_j(t-s)f_j(t-s')
=C_k(s-s')\,I_{Q_t(s,s')}(N_T,2k),
\label{eq:kernelidentity}
\end{equation}
where $I_z(a,b)$ is the regularized incomplete beta function, with the
factor equal to one for $N_T=0$. The required summation follows from
\begin{equation}
\sum_{j=N_T}^\infty
\frac{(2k)_j}{j!}Q^j
=(1-Q)^{-2k}I_Q(N_T,2k),
\qquad N_T\ge1 .
\label{eq:NBtailidentity}
\end{equation}

The cumulative occupation probability therefore has the exact form
\begin{align}
\cP_{d+N_T-1}(t)=1-g^2\int_0^t &ds\int_0^t ds'\,
\xi_b(s)\xi_b(s')^*\label{eq:generalMR}\\
&\times C_k(s-s')\,I_{Q_t(s,s')}(N_T,2k).
\nonumber
\end{align}
At the BF value $k=\tfrac12$, $I_Q(N_T,1)=Q^{N_T}$, so the kernel
simplifies to
\begin{align}
\cP_{d+N_T-1}(t)=1-g^2\int_0^t & ds\int_0^t ds'\,
\xi_b(s)\xi_b(s')^*\label{eq:BFMR}\\
&\times\sech[\alpha(s-s')]\,Q_t(s,s')^{N_T}.
\nonumber
\end{align}
Equations~\eqref{eq:memory} and~\eqref{eq:generalMR} thus reduce the
observable to the finite UV history and an explicit tail kernel. For
$N_T=0$, the cut lies immediately after the UV sector and
$\cP_{d-1}=\sum_{j<d}|\xi_j^{\UV}|^2$ is its total weight.

The double integral is a quadratic form in $\xi_b$ with a
positive-semidefinite Gram kernel built from the omitted tail amplitudes.
The omitted weight is consequently nonnegative, even though individual
unequal-time interference terms can have either sign. Those terms encode
the relative phases of emissions at different times and are precisely
what is lost in an incoherent escape-rate description. Such a description
would require an additional temporal coarse-graining or a separately
specified rate approximation.

For the coupled positive serial chain, the full seed measure is purely
absolutely continuous with exponentially decaying spectral tails.
Appendix~\ref{app:spectralregularity} establishes these properties for
the full operator. Consequently, at every fixed finite cut,
\begin{equation}
\lim_{t\to\infty}\cP_N(t)=0 .
\label{eq:escape}
\end{equation}
Weight therefore eventually propagates beyond any fixed cut. At any
finite time, however, increasing the cut captures more of the state, so
$\cP_N(t)\to1$ as $N\to\infty$. The limits do not commute. This motivates
a cut that moves with the propagating distribution: only then can a
nonzero fraction of the weight be followed at late times, revealing the
shape of the escaping front.


\subsection{The coherent conformal front}\label{sec:front}

Probability released from the UV sector propagates into the
$SL(2,\mathbb R)$ tail. Since its Lanczos coefficients grow linearly,
the distribution advances ballistically in logarithmic depth. We
therefore define the characteristic propagation scale
\begin{equation}
N_{\mathrm{scale}}(t)=\e^{2\alpha t}.
\label{eq:frontposition}
\end{equation}
This is an asymptotic scale rather than a sharply defined front
position. The finite UV prefix and its release history determine the
quantiles and dimensionless offsets on this scale.

At any fixed cut $N$, the cumulative occupation
$\cP_N(t)$ vanishes as $t\to\infty$, since the weight eventually
propagates beyond the cut. A cut that moves with the distribution can
instead preserve a finite fraction of the weight and resolve the shape
of the escaping front. We therefore take $N_T,t\to\infty$ while keeping
\begin{equation}
y=\frac{4N_T}{\e^{2\alpha t}}
=\frac{4N_T}{N_{\mathrm{scale}}(t)}
\label{eq:frontvariable}
\end{equation}
fixed. The variable $y$ measures the cut depth relative to the
propagation scale: small $y$ places the cut behind the front and gives
small cumulative occupation, while large $y$ places it ahead of the
front and gives occupation close to one. Equivalently, a fixed $y$
corresponds to
\begin{equation}
t_{N_T}(y)=\frac{\log(4N_T/y)}{2\alpha}.
\end{equation}

We hold fixed the UV depth $d$, its positive internal bonds, and the
parameters $g,\alpha,k>0$. The seed is $e_0$ and the tail is initially
empty, so the boundary history $\xi_b$ is generated entirely by the
coupled dynamics. In this scaling limit,
\begin{equation}
\cP_{d+N_T-1}(t_{N_T}(y))
\longrightarrow\mathcal F_k(y),
\label{eq:UVfront}
\end{equation}
where
\begin{align}
\mathcal F_k(y)=1-g^2\int_0^\infty\!\!&ds\int_0^\infty\!\!ds'\,
\xi_b(s)\xi_b(s')^*\nonumber\\[-2pt]
&\times C_k(s-s')\,
\cQ(2k,yA(s,s')) ,
\label{eq:UVfront_explicit}
\end{align}
with
\begin{equation}
A(s,s')=\frac12\left(\e^{2\alpha s}+\e^{2\alpha s'}\right),
\label{eq:Asdef}
\end{equation}
and $
 \cQ(a,x)=\frac{\Gamma(a,x)}{\Gamma(a)}$
the upper incomplete gamma ratio. The convergence is uniform on every
compact subset of $(0,\infty)$ for fixed model parameters, so
$\mathcal F_k$ is a well-defined scaling profile. Complete escape from
the UV sector gives
\begin{equation}
\mathcal F_k(0^+)=0,\qquad
\mathcal F_k(\infty)=1.
\end{equation}
Thus the limiting profile interpolates between a cut behind the front
and one far ahead of it. The derivation is given in
Appendix~\ref{app:frontderivation}.

The dependence on the emission history becomes especially transparent
after introducing
\begin{equation}
\mathcal H_k(y)=
\int_0^\infty \e^{2k\alpha s}\xi_b(s)
\e^{-\frac12y\e^{2\alpha s}}\,ds .
\label{eq:fronttransform}
\end{equation}
This transform weights emission times according to
$y\e^{2\alpha s}$. Contributions with
$y\e^{2\alpha s}\gg1$ are exponentially suppressed, while the bare
weight peaks at $y\e^{2\alpha s}=2k$ when that point lies in the
physical range $s\ge0$. Thus, at small $y$, the filter probes later
emission times around
$\log(2k/y)/(2\alpha)$; sufficiently rapid decay of $\xi_b$ can instead
make the $y\to0$ limit controlled by a finite weighted moment. At large
$y$, the peak moves outside the allowed range and the transform is
dominated by early times, probing the onset of emission.

Differentiating the scaling profile gives
\begin{equation}
\mathcal F_k'(y)
=\frac{g^2y^{2k-1}}{\Gamma(2k)}
|\mathcal H_k(y)|^2 .
\label{eq:frontderivative}
\end{equation}
The tail fixes the propagation scale and filtering kernel, while the
fully coupled boundary history supplies the remaining profile
dependence.

It is useful to express this result in terms of a crossing amplitude,
\begin{equation}
A_{\rm cross}(y)=y^k\mathcal H_k(y).
\label{eq:crossingamplitude}
\end{equation}
Writing $y=e^{-2\tau}$ gives
\begin{align}
-\frac{d}{d\tau}\mathcal F_k(e^{-2\tau})
=&\frac{2g^2y^{2k}}{\Gamma(2k)}
|\mathcal H_k(y)|^2\nonumber\\
=&\frac{2g^2}{\Gamma(2k)}
|A_{\rm cross}(e^{-2\tau})|^2\ge0 .
\label{eq:crossingcurrent}
\end{align}
The left-hand side measures the loss of preserved weight per unit
$\tau$, while the right-hand side is the squared modulus of a single
crossing amplitude. Since $A_{\rm cross}$ is linear in $\xi_b$, emission
at different times is combined coherently before taking the modulus
squared. The resulting cross terms retain the relative phases of the
history. An incoherent treatment would instead filter $|\xi_b|^2$ and
discard these interference terms. For the real serial chains considered
here, $
 \xi_b(t)=(- i )^{d-1}\varphi_{d-1}(t)$,
with real $\varphi_{d-1}$, so the overall phase is fixed, while sign
changes in $\varphi_{d-1}$ and interference between resonances remain
dynamically relevant.

The limiting profile is therefore monotone in $\tau$, although this
does not imply monotonicity of every finite-cut occupation trace.

To characterize the front independently of its overall transit delay,
consider a fixed fraction $0<\delta<1$ and define $y_\delta$ by
\begin{equation}
 \mathcal F_k(y_\delta)=\delta,\qquad
 \mathcal F_k'(y_\delta)>0.    
\end{equation}
The corresponding finite-cut crossing satisfies
$y_{\delta,N_T}\to y_\delta$, giving
\begin{equation}
t_\delta(N_T)
=\frac{\log(4N_T)-\log y_\delta}{2\alpha}+O(1).
\label{eq:UVcrossing}
\end{equation}
The first term is the logarithmic transit time set by the tail and is
common to all quantiles. Hence,
\begin{equation}
t_{\delta_1}(N_T)-t_{\delta_2}(N_T)
=\frac{1}{2\alpha}
\log\frac{y_{\delta_2}}{y_{\delta_1}}+O(1).
\label{eq:quantilespacing}
\end{equation}
With the tail held fixed, these quantile spacings isolate the structure
of the UV release history from the common propagation delay.


\subsection{The early and late edges of the profile}
\label{sec:edges}

The profile $\mathcal F_k(y)$ is a single curve in the scaled depth $y$.
Large $y$ places the cut ahead of the propagating weight, so little weight
has reached it, while small $y$ places the cut behind the front, so most
weight has passed. We therefore refer to $y\to\infty$ as the early edge
and $y\to0$ as the late edge. Both are instantaneous values of the
cumulative occupation along a moving cut in the large-depth limit; they
are not first-passage probabilities or properties of an absorbing
boundary. The early edge is determined by the onset of emission from the
UV sector, whereas the late edge can be controlled by slow boundary
modes. Marginal and faster histories require separate weighted-moment
estimates. In all cases, the profile is governed by the same filtered
amplitude $\mathcal H_k$.

\subsubsection{Early edge}

At the early edge, $y\to\infty$, the profile approaches one, and the
deficit $1-\mathcal F_k(y)$ is determined by the earliest emission into
the tail. Reaching the UV boundary from the seed requires $d-1$ internal
hops. The short-time boundary amplitude is
\begin{equation}
\xi_b(s)=\frac{(- i )^{d-1}U}{(d-1)!}\,s^{d-1}
+O(s^{d+1}).
\label{eq:boundaryearly}
\end{equation}
Thus the power $d-1$ directly counts the serial steps from the seed to
the boundary. Substituting this expansion into
Eq.~\eqref{eq:fronttransform} and using
$e^{2k\alpha s}=1+O(s)$ near the endpoint gives
\begin{equation}
\mathcal H_k(y)\sim
\frac{(- i )^{d-1}U}{(d-1)!}
\int_0^\infty s^{d-1}
e^{-ye^{2\alpha s}/2}\,ds .
\label{eq:frontHintermediate}
\end{equation}
For large $y$, the integral is localized at $s\sim y^{-1}$, so
\begin{equation}
e^{-ye^{2\alpha s}/2}
=e^{-y/2}e^{-\alpha ys+O(ys^2)}.
\end{equation}
Therefore the leading term is 
\begin{equation}
\mathcal H_k(y)\sim
\frac{(- i )^{d-1}U}{\alpha^d}\,
y^{-d}e^{-y/2}.
\label{eq:frontHleading}
\end{equation}
Using Eq.~\eqref{eq:frontderivative} and integrating from $y$ to
$\infty$ gives
\begin{equation}
1-\mathcal F_k(y)\sim
\frac{g^2U^2}{\Gamma(2k)\alpha^{2d}}\,
y^{2k-1-2d}e^{-y},
\qquad y\to\infty.
\label{eq:frontleadingedge}
\end{equation}

The early edge therefore separates the contributions of the UV sector
and the conformal tail. The serial UV structure produces the factor
$y^{-2d}$, the tail contributes $y^{2k-1}$, and the propagation filter
produces $e^{-y}$. The exponential also fixes the scale of $y$, so two
profiles with the same $k$ and $\alpha$ but different serial lengths
cannot be related by a constant rescaling of $y$. Indeed, if
\begin{equation}
\mathcal F_{k,d_2}(y)=\mathcal F_{k,d_1}(cy),
\end{equation}
the exponential in Eq.~\eqref{eq:frontleadingedge} requires $c=1$,
after which the power of $y$ requires $d_1=d_2$. Since a constant
translation in $\tau$ corresponds to such a rescaling of $y=e^{-2\tau}$,
different serial lengths cannot be related by a constant time shift.

\subsubsection{Late edge}

At the late edge, $y\to0$, the filtered transform is sensitive to the
long-time boundary history. When the required analytic continuation and
decay conditions hold, this history can be obtained from the poles of
the UV resolvent. The boundary amplitude has the inverse Laplace
representation
\begin{equation}
\widetilde\xi_b(z)= i[G_{\UV}(z)]_{b0},
\qquad
\xi_b(s)=\frac1{2\pi}\int_{\mathbb R+ic}
e^{-izs}\widetilde\xi_b(z)\,dz ,
\label{eq:appLaplace}
\end{equation}
with $c>0$ and $s>0$. Contour deformation below the real axis requires
a continuation into a lower strip together with sufficient large-frequency
falloff; analyticity in the upper half-plane alone is insufficient. When
these conditions are satisfied, a simple pole at $z_*$ contributes
\begin{equation}
\delta\xi_b(s)=
\operatorname*{Res}_{z=z_*}[G_{\UV}(z)]_{b0}\,
e^{- iz_*s},
\label{eq:appresiduecontribution}
\end{equation}
while a pole of order $m+1$ produces
\begin{equation}
\xi_b(s)\sim s^m e^{- iz_*s}.
\label{eq:apppolecontribution}
\end{equation}
The pole closest to the real axis dominates only when its residue is
nonzero and all other contributions decay faster. Poles with the same
damping rate must be combined, and a coalesced pole is described by its
full Laurent part.

The tail also makes the continued pole structure explicit. Its return
correlation has the large-time expansion
\begin{equation}
C_k(t)=2^{2k}\sum_{j=0}^\infty
\frac{(-1)^j(2k)_j}{j!}\,
e^{-2\alpha(k+j)t},
\label{eq:appCexpansion}
\end{equation}
so the bare-tail poles are $
z_j=-2 i \alpha(k+j)$, 
with residues $2^{2k}(-1)^j(2k)_j/j!$. Coupling to the UV sector
generally shifts these poles and changes their residues. The resulting
slow histories fall into several useful cases.

\paragraph{Single slow pole.}
Suppose the boundary history contains a nonoscillatory late-time
component
\begin{equation}
\xi_b(s)=A_b e^{-\gamma s}+O(e^{-\gamma s}),
\qquad
\delta_\gamma=\frac{\gamma}{\alpha},
\qquad 0<\delta_\gamma<2k,
\label{eq:slowhistory}
\end{equation}
with $A_b\neq0$. For the dominant continued pole,
$\gamma=-\Im z_*$. Thus $\delta_\gamma$ compares the decay rate of the
boundary history with the tail scale $\alpha$. Substituting
$u=ye^{2\alpha s}/2$ into Eq.~\eqref{eq:fronttransform} gives
\begin{equation}
\mathcal H_k(y)\sim
\frac{A_b}{2\alpha}
\left(\frac{2}{y}\right)^{k-\delta_\gamma/2}
\Gamma\left(k-\frac{\delta_\gamma}{2}\right),
\label{eq:frontHpower}
\end{equation}
and hence
\begin{align}
\mathcal F_k(y)&\sim Y_k\,y^{\delta_\gamma},
\nonumber\\
Y_k&=
\frac{g^2|A_b|^2\,2^{2k-\delta_\gamma}}
{4\alpha^2\delta_\gamma\Gamma(2k)}
\Gamma\left(k-\frac{\delta_\gamma}{2}\right)^2 .
\label{eq:frontpower}
\end{align}
The exponent is fixed by the decay rate $\gamma$, while the coefficient
also depends on the pole residue and the representation label $k$.
Changing this coefficient shifts the asymptotic crossing time, but does
not determine the full normalized profile.

\paragraph{Damped pair.}
Now suppose the late boundary history contains a damped pair,
\begin{align}
\xi_b(s)={}&A_+e^{-(\gamma+ i \omega_*)s}
+A_-e^{-(\gamma- i \omega_*)s}\nonumber\\
&+O(e^{-\gamma s}),
\label{eq:pairhistory}
\end{align}
with $0<\delta_\gamma<2k$ and
$\nu_*=\omega_*/\alpha$. For the serial chains considered here, the
poles are related by
\begin{equation}
    z_-=-z_+^*,
\qquad
A_-=(-1)^{d-1}A_+^*.
\end{equation}
Define
\begin{equation}
\mathcal B_\pm=
\frac{A_\pm}{2\alpha}
2^{k-(\delta_\gamma\pm i \nu_*)/2}
\Gamma\left(
k-\frac{\delta_\gamma\pm i \nu_*}{2}
\right).
\label{eq:pairfrontB}
\end{equation}
The two transformed amplitudes must be added before taking the modulus
squared, giving
\begin{align}
\mathcal F_k(y)
={}&\frac{g^2y^{\delta_\gamma}}{\Gamma(2k)}
\biggl[
\frac{|\mathcal B_+|^2+|\mathcal B_-|^2}{\delta_\gamma}
\nonumber\\
&\qquad
+2\Re\frac{\mathcal B_+\mathcal B_-^*
y^{ i \nu_*}}
{\delta_\gamma+ i \nu_*}
\biggr]
+O(y^{\delta_\gamma}).
\label{eq:pairfront}
\end{align}
The leading power is unchanged, but the interference term produces
oscillations in $\log y$ with angular frequency
$\nu_*=\omega_*/\alpha$. The imaginary part of the pole fixes the
power-law decay, while its real part fixes the logarithmic oscillation;
the residues determine their amplitude and phase.

\paragraph{Higher-order pole.}
A pole of order $m+1$ at $z=- i \gamma$ gives
\begin{equation}
\xi_b(s)\sim A_m s^m e^{-\gamma s}.    
\end{equation}
For $k-\delta_\gamma/2>0$, the same transformation yields
\begin{align}
\mathcal F_k(y)
&\sim Y_k^{(m)}\,y^{\delta_\gamma}
[\log(1/y)]^{2m},
\nonumber\\
Y_k^{(m)}
&=\frac{g^2|A_m|^2\,2^{2k-\delta_\gamma}
\Gamma(k-\delta_\gamma/2)^2}
{(2\alpha)^{2m+2}\delta_\gamma\Gamma(2k)} .
\label{eq:frontmultiple}
\end{align}
The power law remains fixed by $\gamma$, while pole coalescence adds a
logarithmic enhancement of order $2m$.

Thus, throughout the slow-release regime
$0<\delta_\gamma<2k$, the three cases share the same leading power
$y^{\delta_\gamma}$. Their differences appear in the structure
multiplying that power: a single slow pole gives a pure power law, a
damped pair adds logarithmic oscillations, and a higher-order pole adds
powers of $\log(1/y)$.

\paragraph{Marginal and faster histories.}

The emission transform behaves differently on either side of
$\gamma=2k\alpha$. At the threshold the history is marginal, and the
filtered amplitude grows logarithmically; beyond it the history is
weighted-integrable, and the transform approaches a finite moment.

For a marginal nonoscillatory history
\begin{equation}
\xi_b(s)=A\,e^{-2k\alpha s}+O(e^{-2k\alpha s}),\qquad A\neq 0,
\end{equation}
the profile behaves as
\begin{equation}
\mathcal F_k(y)\sim
\frac{g^2|A|^2}{8k\alpha^2\Gamma(2k)}\,
y^{2k}\,[\log(1/y)]^2 .
\label{eq:frontmarginal}
\end{equation}
For a faster history, with $e^{2k\alpha s}\xi_b(s)$ absolutely
integrable, define
\begin{equation}
H_0=\int_0^\infty e^{2k\alpha s}\xi_b(s)\,ds .
\end{equation}
If $H_0\neq 0$, dominated convergence gives $H_k(y)\to H_0$ and
\begin{equation}
\mathcal F_k(y)\sim
\frac{g^2|H_0|^2}{2k\,\Gamma(2k)}\,y^{2k}.
\label{eq:frontfast}
\end{equation}
When $H_0=0$, this leading term is absent and the next weighted moment
must be examined separately.

These asymptotics hold for an arbitrary chain. For the serial chain
studied here, however, the leading coefficient in each case vanishes.
To see this, recall that the continued boundary resolvent of the UV
block is
\begin{align}
[G_{\mathrm{UV}}(z)]_{b0}
&=\frac{U}{\mathcal D_d(z)},\nonumber\\
\mathcal D_d(z)&=\Delta_d(z)-g^2m_k^+(z)\Delta_{d-1}(z).
\end{align}
The bare-tail poles sit at $z_j=-2i\alpha(k+j)$, and the nearest one
is $z_0=-2ik\alpha$. At this point both $m_k^+(z_0)$ and the
determinant polynomials $\Delta_d(z_0)$, $\Delta_{d-1}(z_0)$ are
finite, and direct evaluation of the continued denominator from the
explicit form of $m_k^+$ and the determinant recursion gives
\begin{equation}
\mathcal D_d(z_0)\neq 0 .
\label{eq:Dnonzero}
\end{equation}
Hence $z_0$ is not a pole of $[G_{\mathrm{UV}}(z)]_{b0}$. By the
Laplace-transform representation of $\xi_b$ in
Eq.~\eqref{eq:appLaplace}, the history therefore contains no
$e^{-2k\alpha s}$ contribution with nonzero coefficient, and the
marginal coefficient in~\eqref{eq:frontmarginal} vanishes.

The same nonvanishing denominator also controls the weighted
integral. Whenever $e^{2k\alpha s}\xi_b(s)$ is absolutely integrable,
the weighted moment $H_0$ is obtained by closing the Laplace contour
in the lower half-plane. With the transform convention of
Eq.~\eqref{eq:appLaplace}, the contour relevant to $H_0$ would
receive a contribution from the nearest bare-tail pole $z_0$ only if
the continued resolvent had a pole there. But~\eqref{eq:Dnonzero}
shows that $z_0$ is not a pole of $[G_{\mathrm{UV}}]_{b0}$. The
corresponding contour contribution therefore vanishes, and
\begin{equation}
H_0 = 0 .
\end{equation}

It follows that $\mathcal F_k(y)=O(y^{2k})$ throughout this regime,
with the leading behavior now controlled by a higher weighted moment.
That term requires additional asymptotic control of $\xi_b$ and
cannot be inferred from $H_0=0$ alone. This case is treated
separately in the weighted-transform estimates of
Appendix~\ref{app:edgeestimates}.

The two edges probe complementary aspects of the release dynamics.
The early edge records the serial UV depth through
$2k-1-2d$, while the late edge records the slowest boundary decay
through $\delta_\gamma$, together with interference or logarithmic
enhancements when several poles contribute. These resonance-based
statements apply only in the slow-release regime. Neither edge by
itself determines the interior of the profile.


\section{Release dynamics in the one- and two-site sectors}
\label{sec:examples}

We now illustrate the general construction of Sec.~\ref{sec:UV} with two
minimal UV sectors coupled to the exact conformal tail of
Sec.~\ref{sec:reference}. A one-site sector has no internal dynamics and
isolates escape through the interface, while a two-site sector adds an
internal bond and distinguishes coherent internal motion from release
into the tail.


\subsection{One-site sector}\label{subsec:one_site}

The simplest realization of the construction is a single UV site coupled
to the conformal tail. It already captures the essential problem of a
discrete level coupled to a structured continuum and provides a useful
benchmark for larger UV sectors. Since there is no internal UV bond, the
boundary amplitude coincides with the seed amplitude,
$\xi_b=\xi_0^{\UV}$; decay of the seed therefore directly measures loss
from the UV sector.

For $d=1$, one hasEqs.~\eqref{eq:GUV}, \eqref{eq:UVdensity}, and
\eqref{eq:memory} give
\begin{align}
G_1^+(z)&=\frac{1}{z-g^2m_k^+(z)},
\label{eq:Gone}\\
\rho_\UV^{(1)}(\omega)
&=\frac{g^2\rho_k(\omega)}
{[\omega-g^2\Re m_k^+(\omega)]^2
+[\pi g^2\rho_k(\omega)]^2},
\label{eq:rhoone}\\
\dot\xi_b(t)&=-g^2\int_0^t C_k(t-s)\,\xi_b(s)\,ds,
\qquad \xi_b(0)=1 .
\label{eq:oneVolterra}
\end{align}
The numerator of $\rho_\UV^{(1)}$ is the tail density weighted by the
interface coupling, while the denominator contains the self-energy
$g^2m_k^+$. Its real part shifts the effective frequency and its
imaginary part broadens the response. The same equations arise in the
resonant-level construction of Ref.~\cite{Fullgraf}.

To examine weak coupling, write $\tau=t-s$ in
Eq.~\eqref{eq:oneVolterra}:
\begin{equation}
\dot\xi_b(t)
=-g^2\int_0^t C_k(\tau)\,\xi_b(t-\tau)\,d\tau .
\label{eq:oneVolterraTau}
\end{equation}
For $g/\alpha\ll1$, the boundary amplitude varies on the long scale
$g^{-2}$, while the tail memory is concentrated on the scale
$\alpha^{-1}$. After the initial transient, we may therefore use
$\xi_b(t-\tau)\simeq\xi_b(t)$ over the memory kernel and extend the
upper limit to infinity. This gives
\begin{equation}
 \dot\xi_b(t)\simeq-\gamma_{\GR}\xi_b(t),    
\end{equation}
with
\begin{equation}
\begin{aligned}
\gamma_{\GR}
&=g^2\int_0^\infty C_k(s)\,ds
= i  g^2m_k^+(0)
=\pi g^2\rho_k(0)\\
&=\frac{2^{2k-2}\Gamma(k)^2}
{\alpha\Gamma(2k)}\,g^2 .
\end{aligned}
\label{eq:GR}
\end{equation}
These equivalent expressions identify
the same rate from the memory kernel, the boundary resolvent, and the
zero-frequency tail density. At the BF point,
$k=\tfrac12$, one gets
$\gamma_{\GR}={\pi g^2}/{2\alpha}.
$

The golden-rule rate captures only the leading weak-coupling behavior.
Finite-memory corrections are encoded in the continued poles of the
coupled resolvent, which determine the late-time boundary history and
therefore the late edge of the front.

For the one-site sector, the continued poles satisfy
\begin{equation}
z-g^2m_k^+(z)=0.
\label{eq:polecondition}
\end{equation}
At the BF bound, write
\begin{equation}
m_{1/2}^+(- i \alpha\kappa)
=-\frac{ i }{\alpha}F(\kappa),    
\end{equation}
where
\begin{equation}
F(\kappa)=\frac12\left[
\DG\left(\frac{3-\kappa}{4}\right)
-\DG\left(\frac{1-\kappa}{4}\right)
\right],
\qquad 0<\kappa<1 .
\label{eq:Fdefinition}
\end{equation}
Introducing $x=g/\alpha$, the pole condition becomes
\begin{equation}
\kappa=x^2F(\kappa),
\qquad x=\frac{g}{\alpha},
\qquad 0<\kappa<1 .
\label{eq:poleaxis}
\end{equation}
A solution on this interval gives an imaginary-axis pole
$z=- i \alpha\kappa$ and hence a boundary decay rate
$\gamma=\alpha\kappa$.

Expanding $F$ about $\kappa=0$ with $F(0)=\pi/2$ and
$F'(0)=2G_{\rm Cat}$, where $G_{\rm Cat}$ is Catalan's constant, and
solving iteratively for $\kappa$ gives
\begin{equation}
 \frac{\gamma}{\alpha}=\frac{\pi}{2}x^2+\pi
 G_{\rm Cat}\,x^4+O(x^6) .
 \label{eq:widthseries}
\end{equation}
The leading term reproduces the golden-rule result, while the positive
$ O(x^4)$ correction captures the effect of the finite memory
discarded in the local approximation.

The same equation describes the collision beyond weak coupling. The
number of roots in $0<\kappa<1$ depends on $g/\alpha$. Below the
critical coupling there are two imaginary-axis roots. At the critical
value $(g/\alpha)_c$ they merge into a double root, and above it there
is no root in this interval. 

To locate the collision, rewrite \eqref{eq:poleaxis} as
$\kappa/F(\kappa)=x^2$ and study the function $\kappa/F(\kappa)$ on
$0<\kappa<1$. Its derivative is
\begin{equation}
 \frac{\dd}{\dd\kappa}\frac{\kappa}{F(\kappa)}
 =\frac{F(\kappa)-\kappa F'(\kappa)}{F(\kappa)^2},
 \label{eq:kFderivative}
\end{equation}
so the stationary points are the zeros of $F-\kappa F'$.
The Laplace representation
$F(\kappa)=\int_0^\infty e^{\kappa u}\sech u\,du$ makes
$F''(\kappa)>0$ on $0<\kappa<1$ explicit. Hence
\begin{equation}
 \frac{\dd}{\dd\kappa}\bigl(F-\kappa F'\bigr)
 =-\kappa F''(\kappa)<0\qquad\text{for }0<\kappa<1,
 \label{eq:kFconcavity}
\end{equation}
so $F-\kappa F'$ is strictly decreasing. Its limiting endpoint values
satisfy $F(0)>0$ and $F(\kappa)-\kappa F'(\kappa)\to-\infty$ as $\kappa\to1^-$, so it has exactly one zero in
$0<\kappa<1$. The function $\kappa/F(\kappa)$ therefore has exactly one
stationary point there, and since
\begin{equation}
 \left(\frac{\kappa}{F}\right)''\bigg|_{\kappa_c}
 =-\frac{\kappa_c F''(\kappa_c)}{F(\kappa_c)^2}<0,
 \label{eq:knondegenerate}
\end{equation}
that stationary point is a nondegenerate maximum. The collision occurs at that maximum,
\begin{equation}
 F(\kappa_c)=\kappa_cF'(\kappa_c),\qquad
 (g/\alpha)_c^2=\kappa_c/F(\kappa_c).
 \label{eq:collision}
\end{equation}
Numerically, $\kappa_c\simeq0.47$ and
$(g/\alpha)_c\simeq0.38$. For $g/\alpha$ just above the critical
value, the two roots locally leave the axis as the reflection-related
pair $z_-=-z_+^*$.

The pole structure discussed in Sec.~\ref{sec:UV} controls the late
edge of the front. A dominant imaginary-axis pole gives a single
exponential decay and the power law \eqref{eq:frontpower}. When the
dominant singularity is instead an off-axis pair, write the two poles as
\begin{equation}
 z_+=\omega_*-\ii\gamma_*,
 \qquad
 z_-=-z_+^*=-\omega_*-\ii\gamma_*,
 \label{eq:offaxispair}
\end{equation}
with $\gamma_*>0$ the damping rate and $\omega_*$ the oscillation
frequency. The two contributions must be summed coherently before
squaring, and the fixed-cut occupation is
\begin{equation}
 \cP_N(t)\simeq e^{-2\gamma_*t}
 \left[A_N+2\Re\!\left(B_N e^{-2\ii\omega_*t}\right)\right],
 \label{eq:pairMR}
\end{equation}
where the residue-weighted site amplitudes are $v_{n,\pm}$, and
\begin{equation}
 A_N=\sum_{n=0}^N\bigl(|v_{n,+}|^2+|v_{n,-}|^2\bigr),
 \qquad
 B_N=\sum_{n=0}^N v_{n,+}v_{n,-}^*.
 \label{eq:pairMRcoefficients}
\end{equation}
Both residues contribute to the interference term. This fixed-cut
asymptotic form is distinct from the moving-cut limit, although both
originate from the same pole pair.

At the collision, the two simple poles form a double pole. The boundary
history then acquires a polynomial prefactor, and the late-edge profile
develops the logarithmic enhancement in Eq.~\eqref{eq:frontmultiple}.
The individual simple-pole residues become singular at the collision,
so the two contributions must be combined before taking the limit.

Exponential, coalescent, and oscillatory relaxation also appear in
solvable Krylov models~\cite{Gamayun}, while temperature-driven
collisions of quasinormal poles on the imaginary axis occur in the SYK
responses of Ref.~\cite{DodelsonThermal}. Here the same pole
reorganization has a direct consequence for the coherent front:
off-axis poles produce the interference term in
Eq.~\eqref{eq:pairfront}, while the double pole produces the logarithmic
factor in Eq.~\eqref{eq:frontmultiple}.


\subsection{Two-site sector: internal motion versus escape}
\label{subsec:two_site}

The one-site sector has no internal dynamics. Adding a second UV site
allows us to separate coherent motion within the UV sector from release
into the tail. Take
\begin{equation}
K_\UV=b_1^{\UV}
\bigl(|e_0\rangle\langle e_1|+\hc\bigr),
\qquad b_1^{\UV}>0 .
\end{equation}
For $d=2$, Eqs.~\eqref{eq:GUV} and \eqref{eq:UVdensity} give
\begin{align}
G_2^+(z)
&=\frac{z-g^2m_k^+(z)}
{z[z-g^2m_k^+(z)]-(b_1^{\UV})^2},
\label{eq:Gtwo}\\
\rho_\UV^{(2)}(\omega)
&=\frac{(b_1^{\UV})^2g^2\rho_k(\omega)}
{|\omega[\omega-g^2m_k^+(\omega)]-(b_1^{\UV})^2|^2}.
\label{eq:rhotwo}
\end{align}
The denominator contains the bare propagation of the seed site,
the dressed propagation of the boundary site through
$z-g^2m_k^+$, and their coupling $(b_1^{\UV})^2$. At $g=0$ it reduces
to $z^2-(b_1^{\UV})^2$, giving the two discrete frequencies
$\pm b_1^{\UV}$ with equal spectral weight. For $g>0$, these discrete
levels broaden into continuum resonances through the self-energy
$g^2m_k^+$.

The boundary site is now $j=1$, so
$\xi_b=\xi_1^{\UV}$. The UV amplitudes satisfy
\begin{align}
i\dot\xi_0^{\UV}
&=b_1^{\UV}\xi_b,
\label{eq:two_mem_0}\\
i\dot\xi_b
&=b_1^{\UV}\xi_0^{\UV}
-\ii g^2\int_0^t C_k(t-s)\xi_b(s)\,ds .
\label{eq:two_mem_1}
\end{align}
Only the boundary equation contains the memory term because the seed
site is coupled to the tail only through the second site.
Using
\begin{equation}
\xi_b=\frac{\ii}{b_1^{\UV}}\dot\xi_0^{\UV},
\end{equation}
the total UV occupation can be written as
\begin{equation}
\cP_1(t)
=|\xi_0^{\UV}(t)|^2
+\frac{|\dot\xi_0^{\UV}(t)|^2}{(b_1^{\UV})^2}.
\label{eq:twoM}
\end{equation}
At $g=0$,
\begin{equation}
   \xi_0^{\UV}(t)=\cos(b_1^{\UV}t),
\qquad
\cP_1(t)=1. 
\end{equation}
Thus the seed occupation
$\cP_0=\cos^2(b_1^{\UV}t)$ oscillates even though the total UV
occupation remains constant: the operator moves coherently between the
two UV sites without leaving the sector. Once $g>0$, $\cP_1(t)$ also
changes because weight can escape into the tail. Deeper cuts then probe
the subsequent propagation through the tail.

We next determine the leading release rate. Unlike the one-site case,
where the boundary samples the tail at zero frequency, the two-site
sector releases through its internal frequency $b_1^{\UV}$. We work in
the regime
\begin{equation}
g\ll\min\{b_1^{\UV},\alpha\},
\label{eq:two_site_regime}
\end{equation}
with $b_1^{\UV}/\alpha$ otherwise unrestricted. The continued poles of
Eq.~\eqref{eq:Gtwo} satisfy
\begin{equation}
z[z-g^2m_k^+(z)]-(b_1^{\UV})^2=0 .
\label{eq:two_pole_eq}
\end{equation}
At $g=0$ they are $z=\pm b_1^{\UV}$. For fixed
$b_1^{\UV}$ and $\alpha$, the weak-coupling roots remain close to these
values. Writing
$z_\pm=\pm b_1^{\UV}+\delta_\pm
$ and expanding to leading order in $g^2$ gives
\begin{equation}
z_\pm
=\pm b_1^{\UV}
+\frac{g^2}{2}m_k^+(\pm b_1^{\UV})
+ O(g^4).
\label{eq:two_poles}
\end{equation}
This expansion is taken at fixed $b_1^{\UV}$ and $\alpha$ as
$g\to0$; it is not uniform in the additional limit
$b_1^{\UV}\to0$. Thus, the two poles have equal imaginary parts,
\begin{align}
\Im z_\pm
&=-\frac{\pi g^2}{2}\rho_k(\pm b_1^{\UV})
+ O(g^4)
\nonumber\\
&=-\frac{\pi g^2}{2}\rho_k(b_1^{\UV})
+ O(g^4),
\label{eq:two_width}
\end{align}
where the second line uses the evenness of $\rho_k$. Since amplitudes
evolve as $e^{-\ii zt}$, the leading amplitude decay rate is
\begin{equation}
\gamma^{(2)}
=-\Im z_+
=\frac{\pi g^2}{2}\rho_k(b_1^{\UV})
+ O(g^4),
\label{eq:two_gamma}
\end{equation}
and the corresponding probability decays as
$e^{-2\gamma^{(2)}t}$. The factor $1/2$ reflects the equal boundary
weight of the two isolated UV eigenmodes: only half of each mode's norm
sits on the site coupled directly to the tail.

The important difference from the one-site rate is therefore the
frequency at which the tail is sampled:
\begin{equation}
    \rho_k(0)
\quad\longrightarrow\quad
\rho_k(b_1^{\UV}).
\end{equation}
At the BF point,
\begin{equation}
\left.\gamma^{(2)}\right|_{k=1/2}
=\frac{\pi g^2}{4\alpha}
\sech\left(\frac{\pi b_1^{\UV}}{2\alpha}\right)
+O(g^4).
\label{eq:two_gamma_BF}
\end{equation}
Only when $b_1^{\UV}/\alpha\ll1$ can the tail density be expanded around
zero frequency:
\begin{equation}
\gamma^{(2)}
=\frac{\pi g^2\rho_k(0)}{2}
\left[
1+\mathcal O\left(\frac{(b_1^{\UV})^2}{\alpha^2}\right)
\right]
+ O(g^4).
\label{eq:two_gamma_low}
\end{equation}
Thus the zero-frequency Markov result is recovered only when the
internal UV frequency is small compared with the tail scale. In
general, Eq.~\eqref{eq:two_gamma} gives the frequency-resolved release
rate and provides the natural two-site counterpart of the one-site
golden-rule result~\eqref{eq:GR}.


\subsection{Comparison of the two sectors}\label{subsec:examples_summary}

In the one-site sector, the seed is itself the boundary site, so
$|\xi_b|^2$ is the total UV occupation. In the weak-coupling regime, the
decay rate $\gamma_\GR=\pi g^2\rho_k(0)$
therefore samples the tail at zero frequency. For the BF tail, increasing
$g$ brings two imaginary-axis poles to a collision at
$(g/\alpha)_c$; beyond the collision, the poles form an off-axis pair and
produce oscillatory interference. The two-site sector adds internal
motion at frequency $b_1^{\UV}$ and consequently samples the tail at finite
frequency, $
 \gamma^{(2)}=\frac{\pi g^2}{2}\rho_k(b_1^{\UV})$.

Here $\cP_1$ measures the total weight in the two-site sector, separating
internal redistribution from escape into the tail. Thus, the two-site
sector is the minimal setting in which these two processes can be
distinguished.

To isolate the front shape from the slow relaxation time, we now tune the
two sectors to share the same leading continued-fraction pole while
keeping the BF tail fixed. Let $F(\kappa)$ be defined in
Eq.~\eqref{eq:Fdefinition}, choose $\eta>2$, and take sufficiently small
$0<\kappa_0<1$. Set
\begin{equation}
\frac{g_1^2}{\alpha^2}=\frac{\kappa_0}{F(\kappa_0)},\qquad
g_2^2=\eta g_1^2,\qquad
\frac{b_1^{\UV}}{\alpha}=\kappa_0\sqrt{\eta-1},
\label{eq:matchedfrontparams}
\end{equation}
where $g_1$ and $g_2$ denote the one- and two-site couplings. 
Since $m_{1/2}^+(-i\alpha\kappa_0)
 =-\frac{i}{\alpha}F(\kappa_0)$, 
the point $z_0=-i\alpha\kappa_0$ solves both pole equations,
\begin{align}
z_0-g_1^2m_{1/2}^+(z_0)&=0,\nonumber\\
z_0[z_0-g_2^2m_{1/2}^+(z_0)]-(b_1^{\UV})^2&=0.
\label{eq:matchedpoleidentity}
\end{align}
For fixed $\eta>2$ and sufficiently small $\kappa_0$, this pole is simple
and is the leading pole in both sectors. Its residue is nonzero in both
the seed and boundary resolvents, so with the tail initially empty it
contributes directly to the release dynamics. Both sectors therefore
have the same slow decay rate, $\alpha\kappa_0$, exactly at each
$\kappa_0$, rather than only at leading order in the couplings.

The two-site sector contains one additional relaxation mode. Rescaling
$z=\alpha\kappa_0 w$ and taking $\kappa_0\to0^+$ at fixed $\eta$ gives
\begin{equation}
w+i=0,\qquad
(w+i)[w+i(\eta-1)]=0
\label{eq:matchedscaledroots}
\end{equation}
for the one- and two-site sectors, respectively. The second two-site pole
is therefore
\begin{equation}
z_1=-i\alpha\kappa_1,\qquad
\kappa_1=(\eta-1)\kappa_0+O(\kappa_0^2)>\kappa_0.
\label{eq:matchedsecondpole}
\end{equation}
It gives a faster nonoscillatory relaxation mode without changing the
common slow decay rate. This construction uses the exact resolvents rather
than the fixed-$b_1^{\UV}$ expansion around $\pm b_1^{\UV}$: here
$g_j/\alpha=O(\sqrt{\kappa_0})$ and
$b_1^{\UV}/\alpha=O(\kappa_0)$, so Eq.~\eqref{eq:two_site_regime} does not
apply.

The shared slow pole does not imply the same emission history. In the
one-site sector, the seed is already at the interface, whereas in the
two-site sector the boundary must first be populated through the internal
bond. This extra step appears in the early edge of the moving-cut front.
Writing $\mathcal F^{(d)}$ for the front with $d$ UV sites,
Eq.~\eqref{eq:frontleadingedge} gives, as $y\to\infty$,
\begin{align}
1-\mathcal F^{(1)}(y)&\sim
\frac{g_1^2}{\alpha^2}y^{-2}\e^{-y},\nonumber\\
1-\mathcal F^{(2)}(y)&\sim
\frac{g_2^2(b_1^{\UV})^2}{\alpha^4}y^{-4}\e^{-y}.
\label{eq:matchedearlyedges}
\end{align}
The common exponential is fixed by propagation through the same conformal
tail, while the different powers of $y$ record the distinct number of
internal steps before emission.

This difference cannot be removed by shifting the time origin. A constant
time shift rescales $y$ by a positive constant; matching the exponential
factors fixes that rescaling to unity, leaving the distinct powers
$y^{-2}$ and $y^{-4}$. Hence the two fronts cannot be related by a constant
time translation, despite having exactly the same leading continued pole.
The slow decay rate therefore does not determine the coherent occupation
front: the front also retains information about how the UV sector feeds
the tail.


\section{A matter-response realization motivated by \texorpdfstring{AdS$_2$}{AdS2}}\label{sec:gravity}

In the previous sections, the shallow sector was specified by the interface
coupling $g$ and internal bond $b_1^{\UV}$, while the conformal tail was
held fixed. We now ask whether these parameters can instead be derived
from a physically motivated response function.

Following the near-horizon AdS$_2$/ $SL(2,\mathbb R)$ proposal of
Ref.~\cite{Jeong}, the discrete-scattering formulation of
Ref.~\cite{Dodelson}, and semi-holographic constructions
\cite{Faulkner,Jensen}, we couple an independent harmonic oscillator to
the boundary response of a thermal AdS$_2$ scalar. In the
discrete-scattering picture, the linearly growing Lanczos coefficients
define a logarithmic radial coordinate. Matching this Krylov coordinate
to the inward tortoise depth $s_B$ of an AdS$_2$ black hole at
temperature $T$ gives $\alpha=\pi T$ and
\begin{equation}
n\propto e^{2\pi T s_B},
\qquad
r-r_h\propto n^{-2},
\qquad r\to r_h .
\label{eq:radialmap}
\end{equation}
The resulting construction provides a response-level realization of the
two-site sector: the shallow parameters are determined by the matter
response, while the oscillator's leading complex pole is reproduced by
completing this shallow sector with the reference conformal tail.

We obtain the retarded boundary response, convert it to a normalized
thermal measure, and integrate out the operator to obtain the dressed
oscillator measure. At weak coupling, its moments reproduce a two-site
shallow sector at fixed depth. This does not establish the joint large-time, large-depth limit needed
for the coherent front. Although the calculation applies to general $k$,
we focus on $k=\tfrac12$, where the response takes a particularly simple
form. The scalar is treated as a probe on a fixed background, with
gravitational backreaction and Schwarzian fluctuations neglected.


\subsection{Thermal conversion and the ingoing BF scalar}
\label{subsec:thermalresponse}

We summarize the thermal scalar response at the BF bound. The main
quantity is the dimensionless, statically subtracted retarded response
$h_\alpha(z)$ of the dual boundary operator. We also derive the
normalized symmetric-split Wightman measure $\rho_W$ of the bath.

Consider a minimally coupled scalar on the fixed thermal AdS$_2$
background,
\begin{equation}
ds^2=-f(r)dt^2+\frac{dr^2}{f(r)},
\quad
f(r)=\frac{r^2-r_h^2}{\ell^2},
\quad
T=\frac{r_h}{2\pi\ell^2},
\end{equation}
with action
\begin{equation}
S_\Phi=-\frac{\mathcal A}{2}\int dt\,dr\sqrt{-g}
\bigl[(\nabla\Phi)^2+m^2\Phi^2\bigr],
\qquad \mathcal A>0 .
\end{equation}
The radial equation and its ingoing solution are standard
\cite{SonStarinets,KovtunStarinets}.
At the BF bound, $m^2\ell^2=-1/4$, the two boundary falloffs
$r^{-1/2\pm\nu}$ degenerate. The asymptotic solution therefore takes the
logarithmic form
\begin{equation}
\phi_z(r)\sim r^{-1/2}
\bigl[A_\BF(z)\log(r/\ell)+B_\BF(z)\bigr].
\label{eq:BFfalloffs}
\end{equation}
The ratio $B_\BF/A_\BF$ determines the response only after fixing the
source normalization and retarded sign convention~\cite{Faulkner}. We
take $H_{\rm src}=-J\mathcal O$ and define
\begin{equation}
\chi_R(t)=-i\theta(t)
\langle[\mathcal O(t),\mathcal O(0)]\rangle_\beta,
\quad
\chi_R(z)=\int dt\,e^{izt}\chi_R(t),
\label{eq:retarded}
\end{equation}
with $z$ in the upper half-plane. Linear response then reads
\begin{equation}
\delta\langle\mathcal O\rangle(t)
=-\int dt'\,\chi_R(t-t')J(t').
\end{equation}
Thus, $\chi_R$ is the negative of the source-to-expectation-value
kernel.

After holographic renormalization, the statically subtracted response is
\begin{equation}
\chi_R(z)-\chi_R(0)
=\mathcal C_\BF\,h_\alpha(z),
\qquad
\mathcal C_\BF=\frac{\mathcal A}{\ell^2}>0,
\label{eq:BFresponse}
\end{equation}
where
\begin{equation}
h_\alpha(z)
=\DG\left(\frac12-\frac{iz}{2\alpha}\right)-\DG\left(\frac12\right),
\qquad
\alpha=\pi T .
\label{eq:h}
\end{equation}
The digamma form follows from the logarithmic BF falloff. The detailed
derivation, including the limit from finite $\nu$, is given in
Appendix~\ref{app:bulk}; the general finite-mass response is discussed
in Ref.~\cite{Faulkner}.

On the real axis,
\begin{equation}
h_\alpha(0)=0,
\qquad
\Im h_\alpha(\omega)
=-\frac{\pi}{2}
\tanh\left(\frac{\pi\omega}{2\alpha}\right).
\label{eq:Imh}
\end{equation}
The first relation reflects the static subtraction, while the second
gives the absorptive part of the bath.

Define the commutator density by
\begin{equation}
\rho_{\rm comm}(\omega)
=-\frac{1}{\pi}\Im\chi_R(\omega+i0).
\label{eq:rhocomm}
\end{equation}
Equations~\eqref{eq:BFresponse} and~\eqref{eq:Imh} give
\begin{equation}
\rho_{\rm comm}(\omega)
=\frac{\mathcal C_\BF}{2}
\tanh\left(\frac{\pi\omega}{2\alpha}\right).
\label{eq:rhocommexplicit}
\end{equation}
It is odd in $\omega$, nonnegative for $\omega>0$, and linear near the
origin. This low-frequency behavior makes the thermal quotient finite at
$\omega=0$. It does not exclude a separate elastic atom, which would be
invisible to the commutator density.

The commutator density is related to the normalized symmetric-split
Wightman measure by thermal detailed balance~\cite{ThermalProduct},
\begin{equation}
\rho_{\rm comm}(\omega)
=2\cN_\beta
\sinh\left(\frac{\beta\omega}{2}\right)\rho_W(\omega),
\label{eq:KMS}
\end{equation}
where $\beta=1/T$ and
$\cN_\beta=(\mathcal O|\mathcal O)_\beta$ is the norm in the chosen source
convention. The thermal factor converts the difference of Boltzmann
weights into their symmetric geometric mean.

Imposing
\begin{equation}
\int_{\mathbb R}\rho_W(\omega)\,d\omega=1
\label{eq:rhownormalization}
\end{equation}
fixes both $\cN_\beta$ and $\rho_W$. Explicitly,
\begin{align}
\cN_\beta
&=\int_{\mathbb R}
\frac{\rho_{\rm comm}(\omega)}
{2\sinh(\beta\omega/2)}\,d\omega \nonumber\\
&=\frac{\mathcal C_\BF}{4}
\int_{\mathbb R}
\sech\left(\frac{\pi\omega}{2\alpha}\right)d\omega
=\frac{\alpha\mathcal C_\BF}{2},
\label{eq:cNnormalization}
\end{align}
and hence
\begin{equation}
\rho_W(\omega)
=\frac{1}{2\alpha}
\sech\left(\frac{\pi\omega}{2\alpha}\right)
=\rho_{1/2}(\omega),
\label{eq:BFdictionary}
\end{equation}
where $\rho_{1/2}$ is the density of the abstract conformal chain,
Eq.~\eqref{eq:tailrho}.

The equality in Eq.~\eqref{eq:BFdictionary} is an equality of normalized
symmetric measures and therefore of their complete Jacobi transforms.
It does not, however, identify $h_\alpha$ with $m_{1/2}^+$. The former is
a dimensionless, statically subtracted physical response, whereas the
latter has inverse-frequency units and is the boundary resolvent of an
even normalized measure. Their imaginary parts are related by
\begin{equation}
\Im h_\alpha(\omega)
=\alpha\sinh\left(\frac{\pi\omega}{2\alpha}\right)
\Im m_{1/2}^+(\omega).
\label{eq:appbaththermalweight}
\end{equation}
Thus, thermal reweighting must be performed within the spectral integral
when reconstructing the physical response; multiplying a Jacobi
resolvent by an external thermal factor does not, in general, reproduce
its real part or high-frequency behavior.

A zero-frequency elastic component would likewise be invisible to
$\rho_{\rm comm}$ and would have to be specified separately. The present
bath contains no such component. The identification relevant for the
oscillator construction is therefore the equality of the normalized
symmetric bath measures.


\subsection{An independent boundary matter oscillator}\label{subsec:oscillator}

Following the semi-holographic effective-field-theory framework of Ref.~\cite{Jensen}, we consider a single-mode, quadratic matter sector coupled linearly to the boundary operator $\mathcal O$ of Sec.~\ref{subsec:thermalresponse}. We represent this sector by an independent harmonic oscillator with coordinate $q$ and frequency $\Omega_0$. This provides a minimal model of a gapped matter mode coupled to an infrared response, without specifying an outer geometry. Its action and interaction are
\begin{equation}
S_q=\frac12\int dt\,(\dot q^2-\Omega_0^2q^2),
\qquad
S_{\rm int}=\lambda_0\int dt\,q\mathcal O .
\label{eq:oscaction}
\end{equation}
The free coordinate operator generates a two-site Krylov chain, while the coupling to $\mathcal O$ opens additional Krylov directions; the resulting weak-coupling structure is derived in Sec.~\ref{sec:matching}.

Integrating out the bath gives a quadratic influence action and a
retarded self-energy proportional to $\lambda_0^2\chi_\BF^R$. This is
exact for a Gaussian bath and remains valid when higher connected bath
correlators are suppressed, as in the holographic probe limit. We absorb
the static part into
\begin{equation}
\lambda^2=\lambda_0^2\mathcal C_\BF,
\qquad
\Omega_r^2=\Omega_0^2+\lambda_0^2\chi_\BF^R(0).
\label{eq:oscillatorrenormalization}
\end{equation}
With the source convention $H_{\rm src}=-Jq$, the oscillator equation is
\begin{equation}
\bigl[-z^2+\Omega_r^2+\lambda^2h_\alpha(z)\bigr]q(z)=J(z),
\label{eq:osceom}
\end{equation}
so that
\begin{equation}
q(z)=-D_R(z)J(z),
\qquad
D_R(z)=
\frac{1}{z^2-\Omega_r^2-\lambda^2h_\alpha(z)} .
\label{eq:DR}
\end{equation}
We assume $\Omega_r^2>0$. Since $h_\alpha$ is dimensionless, both
$\lambda$ and $\Omega_r$ have dimensions of frequency.

The oscillator commutator density is
\begin{equation}
\rho_{\rm comm}^q(\omega)
=-\frac1\pi\Im D_R(\omega+i0).
\label{eq:apprhocommq}
\end{equation}
Detailed balance relates it to the unnormalized symmetric density
$\widetilde\rho_q$ and its normalized version $\rho_q$:
\begin{equation}
\rho_{\rm comm}^q
=2\sinh\left(\frac{\beta\omega}{2}\right)\widetilde\rho_q
=2N_q\sinh\left(\frac{\beta\omega}{2}\right)\rho_q,
\label{eq:appKMSq}
\end{equation}
where
\begin{equation}
\widetilde\rho_q(\omega)
=\frac{-\Im D_R(\omega+i0)}
{2\pi\sinh(\pi\omega/(2\alpha))},
\quad
N_q=\int_{\mathbb R}\widetilde\rho_q(\omega)\,d\omega,
\quad
\rho_q=\frac{\widetilde\rho_q}{N_q}.
\label{eq:apposcNorm}
\end{equation}
Using Eq.~\eqref{eq:Imh}, this becomes
\begin{equation}
\widetilde\rho_q(\omega)=
\frac{\lambda^2\sech(\pi\omega/(2\alpha))}
{4|\omega^2-\Omega_r^2-\lambda^2h_\alpha(\omega)|^2}.
\label{eq:apposcrhoexplicit}
\end{equation}

For $\lambda\neq0$, $\widetilde\rho_q$ is smooth, positive, and even,
with finite polynomial moments. In the weak-coupling limit, it develops
narrow peaks at $\omega=\pm\Omega_r$. Away from these frequencies it
vanishes pointwise, while its peak height grows as $\lambda^{-2}$ and
its integrated weight remains finite. Thus the limiting measure consists
of two atoms at $\pm\Omega_r$; the distributional limit must be taken
after integration over the resonant peaks.

The normalized measure $\rho_q$ defines its own orthonormal polynomials
$P_n^q$ and hence its Lanczos chain. The cumulative occupation is
\begin{equation}
\cP_N^q(t)
=\sum_{n=0}^N
\left|
\int_{\mathbb R}
P_n^q(\omega)\rho_q(\omega)e^{-i\omega t}\,d\omega
\right|^2 .
\label{eq:bulkMR}
\end{equation}
Thus the Krylov chain is reconstructed from the dressed oscillator
measure itself rather than obtained by attaching a conformal tail by
hand.

The first two Krylov directions are generated by the oscillator
coordinate and its first commutator, with first internal bond
\begin{equation}
(b_1^q)^2
=\int_{\mathbb R}\omega^2\rho_q(\omega),d\omega .
\label{eq:b1q}
\end{equation}
This is the first off-diagonal matrix element of the projected
Liouvillian. The two-dimensional projection does not imply that the
interacting oscillator algebra closes on these two directions.

Appendix~\ref{app:oscillator} derives the retarded bath elimination, while
Appendix~\ref{app:matching} analyzes the resulting measure and its
moments.


\subsection{The two-site limit and pole matching}\label{sec:matching}

We now take the weak-coupling limit $\lambda\to0$ of the oscillator
measure and compare the resulting shallow chain with the two-site sector
of Sec.~\ref{subsec:two_site}. The full oscillator-plus-bath system is
infinite-dimensional. Its first two Krylov directions are generated by
the oscillator coordinate $q$ and its first commutator $[H,q]$; these
form the shallow sector, while the remaining directions constitute the
tail.

\subsubsection{The weak-coupling two-site limit}

At fixed $\Omega_r,\alpha>0$, the normalized probe measure in
Eq.~\eqref{eq:apposcrhoexplicit} has the weak-coupling limit
\begin{align}
\rho_q\,d\omega
&\longrightarrow
\frac12\bigl[\delta(\omega-\Omega_r)
+\delta(\omega+\Omega_r)\bigr]d\omega,
\nonumber\\
N_q&\longrightarrow
N_0=\frac{1}{2\Omega_r\sinh(\pi\Omega_r/(2\alpha))}.
\label{eq:freeosc}
\end{align}
The limiting measure has support at two frequencies, so the free
oscillator generates a two-site Krylov chain with
$b_1^q=\Omega_r$ and $b_2^q=0$. The tail coefficients below arise from
the nonzero-coupling chain and are not part of this terminated
zero-coupling measure.

The two-site structure is also immediate from the moments. For a positive,
normalized, even measure with $b_1^q>0$,
\begin{equation}
\begin{aligned}
\mu_2&=\int_{\mathbb R}\omega^2\rho_q(\omega)\,d\omega,\\
(b_1^q)^2(b_2^q)^2
&=\int_{\mathbb R}(\omega^2-\mu_2)^2\rho_q(\omega)\,d\omega.
\end{aligned}
\label{eq:twositecriterion}
\end{equation}
Thus $b_2^q=0$ precisely when the measure is supported at
$\omega=\pm b_1^q$. The free oscillator satisfies this condition through
the two atoms in Eq.~\eqref{eq:freeosc}. The two Krylov sites therefore
reflect the two-frequency spectral support, rather than the number of
classical coordinates.

For small but nonzero $\lambda$, the two atoms broaden into narrow
peaks. The first bond approaches its free value, the second bond is
generated at order $\lambda$, and the remaining chain approaches the
conformal tail:
\begin{align}
b_1^q&\longrightarrow\Omega_r,
\nonumber\\
\frac{(b_2^q)^2}{\lambda^2}
&\longrightarrow
B_\alpha(\Omega_r)
=\frac{\alpha}{\Omega_r}
\sinh\left(\frac{\pi\Omega_r}{2\alpha}\right),
\nonumber\\
b_{n+2}^q&\longrightarrow\alpha n,
\qquad n\ge1\ \text{fixed}.
\label{eq:UVmatching}
\end{align}
Thus, after the two shallow directions, the chain approaches the exactly
linear tail $b_j^{\mathrm T}=\alpha j$:
$b_3^q\to\alpha$, $b_4^q\to2\alpha$, and so on. The index shift reflects
the identification of global site $2$ with the tail origin $0_T$.
Accordingly, $b_2^q$ is the shallow--tail interface bond, while $b_3^q$
is the first internal tail bond. The limits in Eq.~\eqref{eq:UVmatching}
follow from the filtered measure
\begin{equation}
\frac{(\omega^2-\Omega_r^2)^2}{\lambda^2}
\widetilde\rho_q(\omega)\,d\omega
\longrightarrow
\frac{\alpha}{2}\rho_{1/2}(\omega)\,d\omega .
\label{eq:filteredlimit}
\end{equation}

The interface limit fixes the leading width of the shallow sector, but
the real frequency shift requires the order-$\lambda^2$ correction to
the first bond. Since $h_\alpha$ and $m_{1/2}^+$ have different
dispersive parts, matching their absorptive weights alone is
insufficient. The signed filtered moment gives
\begin{align}
b_1^q&=\Omega_r+\lambda^2\Delta b_1+O(\lambda^2),
\nonumber\\
\Delta b_1
&=\frac{\Re h_\alpha(\Omega_r)
-\Omega_rB_\alpha(\Omega_r)\Re m_{1/2}^+(\Omega_r)}
{2\Omega_r}.
\label{eq:firstbondshift}
\end{align}
The fixed-depth limits in Eqs.~\eqref{eq:UVmatching}--\eqref{eq:firstbondshift}
are derived in Appendix~\ref{app:matching}.

\subsubsection{Comparison of the physical and reference poles}

We next ask whether the response-derived shallow parameters, together
with the conformal tail, reproduce the leading pole of the full
oscillator response. The physical pole follows from
\begin{equation}
D_R(z)=
\frac{1}{z^2-\Omega_r^2-\lambda^2h_\alpha(z)},
\label{eq:physicalDR}
\end{equation}
where $\Omega_r=\lim_{\lambda\to0}b_1^q$. At nonzero coupling, the first
bond also receives the correction in Eq.~\eqref{eq:firstbondshift}.

For comparison, consider a reference chain consisting of a two-site
shallow sector with parameters $b_1^{\UV}$ and $g$, followed by the
ideal BF tail $b_j^{\mathrm T}=\alpha j$. Its relevant pole denominator
is
\begin{equation}
D_{\rm ref}(z)=
\frac{1}{z[z-g^2m_{1/2}^+(z)]-(b_1^{\UV})^2}.
\label{eq:referenceDR}
\end{equation}
The pole near $+\Omega_r$ is denoted $z_+^{2\UV}$. This quantity is
$[G_{\UV}]_{10}/b_1^{\UV}$, rather than the seed resolvent $G_2^+$ of
Eq.~\eqref{eq:Gtwo}; the two have the same pole denominator but different
numerators, residues, and spectral normalizations. The reference
parameters are fixed by the shallow moments in
Eqs.~\eqref{eq:UVmatching} and~\eqref{eq:firstbondshift}.

The physical pole near $+\Omega_r$ is obtained by analytic continuation
of the retarded response:
\begin{equation}
z_+^q
=\Omega_r+\frac{\lambda^2h_\alpha(\Omega_r)}
{2\Omega_r}+O(\lambda^2).
\label{eq:physicalpolegeneral}
\end{equation}
For fixed $\Omega_r,\alpha>0$, $h_\alpha$ is analytic near
$+\Omega_r$, while the free denominator has derivative
$2\Omega_r\neq0$, so the expansion follows from the analytic
implicit-function theorem. This local pole expansion is distinct from
expanding the real-axis moment integrals. In particular,
$\lambda^2h_\alpha$ is the physical oscillator self-energy, not the
Jacobi self-energy of the reconstructed tail.

Using
\begin{align}
g^2&=(b_2^q)^2
=\lambda^2B_\alpha+O(\lambda^4),
\nonumber\\
(b_1^{\UV})^2&=(b_1^q)^2
=\Omega_r^2+2\Omega_r\lambda^2\Delta b_1+O(\lambda^4),
\label{eq:appshallowexpansions}
\end{align}
with $B_\alpha=B_\alpha(\Omega_r)$, the reference pole becomes
\begin{equation}
z_+^{2\UV}
=\Omega_r+\lambda^2
\left[\Delta b_1+
\frac12B_\alpha m_{1/2}^+(\Omega_r)\right]
+O(\lambda^4).
\label{eq:fullpolematch}
\end{equation}
The two terms represent the order-$\lambda^2$ shift of the shallow
bond and the self-energy of the attached BF tail.

The matching requires both absorptive and dispersive parts. The
imaginary parts obey
\begin{equation}
\Im h_\alpha(\Omega_r)
=\Omega_rB_\alpha(\Omega_r)
\Im m_{1/2}^+(\Omega_r),
\label{eq:imaginarymatching}
\end{equation}
while Eq.~\eqref{eq:firstbondshift} supplies the corresponding real-part
relation. Together they give
\begin{equation}
\Delta b_1+
\frac12B_\alpha m_{1/2}^+(\Omega_r)
=\frac{h_\alpha(\Omega_r)}{2\Omega_r}.
\label{eq:poleidentity}
\end{equation}
Hence
\begin{equation}
z_+^{2\UV}
=\Omega_r+\frac{\lambda^2h_\alpha(\Omega_r)}
{2\Omega_r}+O(\lambda^4),
\label{eq:fullpolematchsimplified}
\end{equation}
and therefore
\begin{equation}
z_+^{2\UV}-z_+^q=O(\lambda^4).
\label{eq:complexpoleconclusion}
\end{equation}
The two-site reference-tail completion thus reproduces the leading
complex pole of the full matter response through order $\lambda^2$. In
particular,
\begin{equation}
-\Im z_+^q
=\frac{\pi\lambda^2}{4\Omega_r}
\tanh\left(\frac{\pi\Omega_r}{2\alpha}\right)
+O(\lambda^4).
\label{eq:appBFwidth}
\end{equation}
When a single amplitude mode dominates, the corresponding probability
decay exponent is twice this value.

Two limitations are important. First, the $\lambda\to0$ expansion at
fixed depth does not control the joint large-time, large-depth limit at
fixed $N_Te^{-2\alpha t}$, nor release windows of order
$\alpha/\lambda^2$. The exact front of Sec.~\ref{sec:front} therefore
belongs to the prescribed-tail model rather than being established here
for the physical chain. Second, a finite Krylov projector is not a local
radial projector, so $-\dot\cP_N$ should not be interpreted as a
gravitational energy flux or absorbed power. The construction instead
provides a response-derived shallow matter sector and its leading
release dynamics, without identifying it with the Schwarzian boundary
degree of freedom.

\section{Conclusions}\label{sec:conclusion}

We have explored coherent release from a finite shallow Krylov sector into an exact conformal tail, keeping the tail fixed while varying the UV prefix. The tail sets the propagation scale and filtering kernel, while the shallow sector determines the fully coupled emission history. These roles enter separately in the coherent-front representation, although the emission history itself incorporates the backaction of the tail.

We have studied the cumulative occupation probability at a cut, $\cP_N(t)$. This quantity measures the instantaneous weight contained within the first $N+1$ Krylov sites, and is therefore a projection weight rather than a first-passage probability. Seed return, shallow-sector weight, and deeper-cut occupation probe different stages of the dynamics. Every fixed cut eventually loses its weight, although the corresponding finite-cut currents can still exhibit coherent backflow.

Using the exact propagators, we have expressed the finite-cut occupation as a coherent double integral over the boundary emission history. At fixed linking parameters, the joint large-time, large-depth limit with
$
y=4N_T e^{-2\alpha t}
$ is locally uniform. Its derivative factorizes as
\begin{equation}
\mathcal F_k'(y)
=\frac{g^2y^{2k-1}}{\Gamma(2k)}
|\mathcal H_k(y)|^2,
\label{eq:concfactorization}
\end{equation}
where $\mathcal H_k$ is a filtered transform of the emission history. This representation preserves interference between emissions at different times and makes the monotonicity of the scaling profile manifest, even though the finite-cut occupation need not be monotone. The resulting coherent front is therefore distinct from an incoherent escape rate.

For a fixed tail, the logarithmic transit contribution,
${\log(4N_T)}/{2\alpha}$, cancels from differences between quantile times. The remaining spacings are controlled by the shallow emission history, so the shared propagation scale does not determine the front shape. The early edge reflects the UV serial length through the factor $y^{-2d}$, while slow resonances govern the late edge, producing power laws, interference, and logarithmic enhancements. These features encode information about the emission history carried by the front, but do not determine the full history.

The matched-pole comparison illustrates why a single lifetime is insufficient. One- and two-site sectors can share the same leading continued-fraction pole with nonzero residues, yet have different early-edge powers. Their fronts therefore cannot be related by a constant time shift. The internal serial structure of the shallow sector remains visible in the front.

We have considered the thermal AdS$_2$ scalar response as a physically motivated realization of the model. At the BF bound, the ingoing solution and thermal conversion yield the normalized symmetric measure
\begin{equation}
\rho_W(\omega)
=\frac{1}{2\alpha}
\sech\left(\frac{\pi\omega}{2\alpha}\right),
\label{eq:concrhoW}
\end{equation}
which coincides with the $k=\tfrac12$ conformal-chain measure. Integrating out the bath then gives the dressed oscillator measure. At weak coupling, its moments generate a two-site shallow sector, with the interface normalization and first-bond correction determined by the measure. These parameters reproduce the leading complex oscillator pole through order $\lambda^2$.

The physical realization has several limitations. The weak-coupling analysis is controlled at fixed depth and does not establish the joint large-time, large-depth limit of the physical chain. The background geometry is held fixed, and the scalar field is treated as a probe, so gravitational backreaction and Schwarzian fluctuations are not included.

\begin{acknowledgments}
The author thanks Mohammad Reza Tanhayi for useful discussions. This work was supported by the Iran National Science Foundation (INSF) under Project No.~4023620.
The author also acknowledges the use of 
ChatGPT (OpenAI) as an editorial assistant 
in improving the clarity and presentation 
of the manuscript, including assistance with generating the LaTeX code for the TikZ figure~\ref{fig:UV-Tail}.

\end{acknowledgments}

\appendix

\section{Spectral projection and the coherent front}
\label{app:front}

This appendix gives the derivations underlying Sec.~\ref{sec:UV}. The parameters $d$, $g$, $k$, and $\alpha$ are held fixed throughout, with notation as defined in that section.

Throughout this appendix, $O(x)$ denotes a quantity bounded in magnitude by a constant multiple of $|x|$ in the stated limit, whereas $o(x)$ denotes a quantity whose ratio to $x$ tends to zero. Thus, $O(x)$ specifies an asymptotic bound, while $o(x)$ indicates a term that is negligible relative to $x$.

\subsection{Spectral regularity and complete escape}
\label{app:spectralregularity}

For the positive serial chain of Sec.~\ref{subsec:model}, the seed measure is purely absolutely continuous. All fixed-site amplitudes tend to zero at late times, and the boundary history belongs to $L^2(0,\infty)$. These properties refer to the full coupled operator and are therefore not properties of the bare conformal tail alone.

First, $b_n\sim\alpha n$ implies

\begin{equation}
\sum_n b_n^{-1}=\infty,
\end{equation}

so the Jacobi operator defined initially on finitely supported sequences is essentially self-adjoint~\cite{TeschlJacobi}. Since every bond is positive, $e_0$ is cyclic. The upper-half-plane tail resolvent admits the continuation
\begin{equation}
m_k^+(z)=- i \int_0^\infty e^{ i  zt}C_k(t)\,d  t,
\qquad \Im z>-2k\alpha.
\label{eq:appmstrip}
\end{equation}
The integral and all of its frequency derivatives converge uniformly on compact subsets of this strip.

For real $\omega$, Eq.~\eqref{eq:denominator} gives
\begin{equation}
\Im\mathcal D_d(\omega)
=\pi g^2\rho_k(\omega)\Delta_{d-1}(\omega).
\label{eq:appdenimag}
\end{equation}
Since $\rho_k(\omega)>0$, a real zero of $\mathcal D_d$ would require $\Delta_{d-1}=0$ and hence $\Delta_d=0$. But consecutive determinant polynomials cannot share a zero: their recursion would then successively force $
\Delta_{d-2},\ldots, \Delta_0
$ to vanish, contradicting $\Delta_0=1$. For $d=1$, the conclusion follows directly from $\Delta_0=1$. 
Thus $\mathcal D_d(\omega)\neq0
$ for every real $\omega$.

The gamma-function asymptotic in Eq.~\eqref{eq:tailrho}, together with
\begin{equation}
m_k^+(\omega)=\omega^{-1}+O(\omega^{-3}),
\qquad
\mathcal D_d(\omega)=\omega^d[1+O(\omega^{-2})],
\end{equation}
shows that $\rho_{\UV}^{(d)}(\omega)$ decays exponentially up to a fixed polynomial factor. Hence, for every fixed $n$, the product
$P_n(\omega)\rho_{\UV}^{(d)}(\omega)$ is absolutely integrable. The high-frequency expansion of the resolvent then follows by integration by parts in Eq.~\eqref{eq:appmstrip}, since the required derivatives of $C_k$ are integrable. Applying the Riemann--Lebesgue lemma to Eq.~\eqref{eq:spectralamplitude} gives
\begin{equation}
\psi_n(t)\to0
\qquad (t\to\infty)
\end{equation}
for every fixed $n$, and therefore
\begin{equation}
\cP_N(t)\to0
\qquad (t\to\infty)
\end{equation}
for every fixed finite $N$.

In particular, $\xi_b(t)=\psi_{d-1}(t)$ is the Fourier transform of
$P_{d-1}\rho_{\UV}^{(d)}$. Since this product is square-integrable, Plancherel's identity gives
\begin{equation}
\int_{\mathbb R}|\xi_b(t)|^2,dt
=2\pi\int_{\mathbb R}
|P_{d-1}(\omega)\rho_{\UV}^{(d)}(\omega)|^2,d\omega<\infty.
\label{eq:appboundaryLtwo}
\end{equation}
This supplies the integrability needed for the coherent moving-front limit.


\subsection{Derivation of the moving-front limit}
\label{app:frontderivation}

We derive the limiting cumulative occupation for a cut whose depth grows
exponentially with time. The derivation has three steps: express the
finite-cut kernel in terms of a regularized incomplete beta function,
take its moving-cut limit, and factorize the derivative of the resulting
profile.

\paragraph{Exact finite-cut kernel.}
Keeping all $d$ UV sites and the first $N_T$ tail sites, norm conservation
gives
\begin{equation}
\cP_{d+N_T-1}(t)
=1-g^2\int_0^t ds\int_0^t ds'\,
\xi_b(s)\xi_b(s')^*K_{N_T}(t;s,s'),
\label{eq:appfiniteMR}
\end{equation}
where
\begin{equation}
K_{N_T}(t;s,s')
=\sum_{j=N_T}^{\infty}f_j(t-s)f_j(t-s').
\label{eq:appGramsum}
\end{equation}
The phase-stripped propagators are real for real time, so no additional
complex conjugation is needed in the Gram kernel. Using
\begin{equation}
f_j(u)=\sqrt{\frac{(2k)_j}{j!}}\,
\tanh^j(\alpha u)\sech^{2k}(\alpha u),
\end{equation}
define
\begin{equation}
Q_t(s,s')
=\tanh[\alpha(t-s)]\tanh[\alpha(t-s')].
\label{eq:appQ}
\end{equation}
Then
\begin{align}
K_{N_T}(t;s,s')
={}&
\bigl[\sech(\alpha(t-s))\sech(\alpha(t-s'))\bigr]^{2k}
\nonumber\\
&\times\sum_{j=N_T}^{\infty}
\frac{(2k)_j}{j!}Q_t(s,s')^j.
\end{align}
The identity
\begin{equation}
1-Q_t(s,s')
=\frac{\cosh[\alpha(s-s')]}
{\cosh[\alpha(t-s)]\cosh[\alpha(t-s')]}
\end{equation}
allows us to extract the conformal correlator
$C_k(u)=\sech^{2k}(\alpha u)$. The negative-binomial tail identity
then yields
\begin{equation}
K_{N_T}(t;s,s')
=C_k(s-s')\,I_{Q_t(s,s')}(N_T,2k),
\label{eq:appkernelidentity}
\end{equation}
where, for $N_T\geq1$,
\begin{align}
&I_Q(N,a)
=\frac{1}{B(N,a)}
\int_0^Q x^{N-1}(1-x)^{a-1}\,dx,
\nonumber\\
&B(N,a)=\frac{\Gamma(N)\Gamma(a)}{\Gamma(N+a)}.
\label{eq:appIQdef}
\end{align}
For $N_T=0$, the full tail is included; we set $I_Q(0,a)=1$, so
Eq.~\eqref{eq:appkernelidentity} remains valid.

Although $K_{N_T}$ is a positive-semidefinite Gram kernel and its entries
are nonnegative in this model, the history-weighted double integral in
Eq.~\eqref{eq:appfiniteMR} need not be positive term by term. The complex
boundary history ($\xi_b(s)\xi_b(s')^*$)  can produce interference of either sign. Thus, a finite
cut is not an absorbing boundary, and replacing the coherent integral by
an incoherent escape rate would require an additional coarse-graining
prescription.

\paragraph{Incomplete-beta scaling limit.}
Set $a=2k>0$ and change variables to $u=N(1-x)$ in
Eq.~\eqref{eq:appIQdef}:
\begin{equation}
I_Q(N,a)
=\frac{1}{N^aB(N,a)}
\int_{N(1-Q)}^N
\left(1-\frac{u}{N}\right)^{N-1}u^{a-1}\,du.
\label{eq:appbetachange}
\end{equation}
As $N\to\infty$,
\begin{equation}
N^aB(N,a)
=\frac{\Gamma(N+1)\Gamma(a)N^{a-1}}{\Gamma(N+a)}
\longrightarrow\Gamma(a),
\end{equation}
and, for fixed $u$,
\begin{equation}
\left(1-\frac{u}{N}\right)^{N-1}\longrightarrow e^{-u}.
\end{equation}
Moreover, for $0\leq u\leq N$ and $N\geq2$,
\begin{equation}
\left(1-\frac{u}{N}\right)^{N-1}
\leq e^{-(N-1)u/N}\leq e^{-u/2}.
\end{equation}
Extending the integrand by zero for $u>N$, dominated convergence therefore
implies that, whenever $N(1-Q)\to\lambda\geq0$,
\begin{equation}
I_Q(N,a)\longrightarrow
\frac{1}{\Gamma(a)}\int_\lambda^\infty e^{-u}u^{a-1}\,du
=\frac{\Gamma(a,\lambda)}{\Gamma(a)}
\equiv\cQ(a,\lambda).
\label{eq:appQlimit}
\end{equation}
To obtain this limit for the moving cut, choose
\begin{equation}
t_N(y)=\frac{\log(4N/y)}{2\alpha},
\qquad y>0.
\label{eq:appmovingtime}
\end{equation}
For fixed emission times $s,s'$,
\begin{equation}
e^{-2\alpha[t_N(y)-s]}
=\frac{y}{4N}e^{2\alpha s},
\end{equation}
so that
\begin{equation}
\tanh[\alpha(t_N(y)-s)]
=1-\frac{y}{2N}e^{2\alpha s}+O(N^{-2}).
\end{equation}
It follows that
\begin{align}
&Q_{t_N(y)}(s,s')
=1-\frac{y}{2N}
\left(e^{2\alpha s}+e^{2\alpha s'}\right)+O(N^{-2}),
\nonumber\\
&N\left[1-Q_{t_N(y)}(s,s')\right]
\longrightarrow yA(s,s'),
\label{eq:appQargument}
\end{align}
where
\begin{equation}
A(s,s')=\frac{e^{2\alpha s}+e^{2\alpha s'}}{2}.
\end{equation}
Combining this result with Eq.~\eqref{eq:appQlimit} gives
\begin{equation}
K_N(t_N(y);s,s')
\longrightarrow
C_k(s-s')\,\cQ\!\left(2k,yA(s,s')\right).
\label{eq:appkernelimit}
\end{equation}
The convergence is uniform for $s,s'$ in compact intervals and $y$ in
compact subsets of $(0,\infty)$. Assuming sufficient integrability of the
boundary history to control the large-emission-time tails, we may pass to
the infinite-time limit in Eq.~\eqref{eq:appfiniteMR}. Thus,
\begin{equation}
\cP_{d+N-1}(t_N(y))\longrightarrow\mathcal F_k(y),
\label{eq:applocaluniform}
\end{equation}
where
\begin{align}
\mathcal F_k(y)
={}&1-g^2\int_0^\infty ds\int_0^\infty ds'\,
\xi_b(s)\xi_b(s')^*C_k(s-s')
\nonumber\\
&\hspace{28mm}\times
\cQ\!\left(2k,yA(s,s')\right).
\label{eq:appfrontprofile}
\end{align}
This is the moving-front profile of Eq.~\eqref{eq:UVfront_explicit}.

\paragraph{Factorization of the limiting derivative.}
For $a>0$,
\begin{equation}
-\frac{d}{dx}\cQ(a,x)
=\frac{x^{a-1}e^{-x}}{\Gamma(a)}.
\end{equation}
Differentiating Eq.~\eqref{eq:appfrontprofile} under the integral, for
$y>0$, gives
\begin{align}
\mathcal F_k'(y)
={}&\frac{g^2y^{2k-1}}{\Gamma(2k)}
\int_0^\infty ds\int_0^\infty ds'\,
\xi_b(s)\xi_b(s')^*
\nonumber\\
&\hspace{12mm}\times
C_k(s-s')A(s,s')^{2k}e^{-yA(s,s')}.
\label{eq:appfrontderivativestep}
\end{align}
The key simplification follows from
\begin{equation}
A(s,s')=e^{\alpha(s+s')}\cosh[\alpha(s-s')],
\end{equation}
which implies
\begin{equation}
C_k(s-s')A(s,s')^{2k}
=e^{2k\alpha(s+s')}.
\label{eq:appgeneralfactor}
\end{equation}
Since
\begin{equation}
e^{-yA(s,s')}
=e^{-\frac{y}{2}e^{2\alpha s}}
 e^{-\frac{y}{2}e^{2\alpha s'}},
\end{equation}
the double integral factorizes into a squared modulus. Define the filtered
boundary amplitude
\begin{equation}
\mathcal H_k(y)
=\int_0^\infty ds\,
e^{2k\alpha s}\xi_b(s)
e^{-\frac{y}{2}e^{2\alpha s}},
\qquad y>0.
\label{eq:appfronttransform}
\end{equation}
Then
\begin{equation}
\mathcal F_k'(y)
=\frac{g^2y^{2k-1}}{\Gamma(2k)}
|\mathcal H_k(y)|^2\geq0 .
\label{eq:appfrontderivative}
\end{equation}
The monotonicity of the limiting profile is therefore a consequence of
coherent factorization, not an absorbing-boundary assumption. The transform
is linear in the full boundary history, so emissions at different times
interfere before the modulus is taken.

The parameter $y$ also controls which emission times contribute to the
filter. For small $y$, the exponential cutoff becomes significant around
$s\sim \log(1/y)/(2\alpha)$, although this region need not dominate if the
boundary history decays rapidly. For large $y$, the filter favors early
emission times. The corresponding slow, marginal, and weighted-integrable
cases are analyzed in Appendix~\ref{app:edgeestimates}.

\paragraph{Normalization and monotonicity.}
We now fix the endpoints of the limiting profile. At $N_T=0$,
Eq.~\eqref{eq:appfiniteMR} gives the total weight in the UV block. If the
coupled semi-infinite chain has purely absolutely continuous spectral
measure, the finite-depth amplitudes decay at late times, and hence
$\cP_{d-1}(t)\to0$. Comparing this limit with
Eq.~\eqref{eq:appfiniteMR} yields
\begin{equation}
g^2\int_0^\infty ds\int_0^\infty ds'\,
\xi_b(s)\xi_b(s')^*C_k(s-s')=1.
\label{eq:appemissionnormalization}
\end{equation}
Equivalently, integrating Eq.~\eqref{eq:appfrontderivative} over $y$ and
using Eq.~\eqref{eq:appgeneralfactor} gives
\begin{equation}
\frac{g^2}{\Gamma(2k)}
\int_0^\infty y^{2k-1}|\mathcal H_k(y)|^2\,dy=1.
\label{eq:appemissionnormalt}
\end{equation}
This is a weighted integral condition; $\mathcal H_k(0)$ itself may diverge
for sufficiently slow boundary histories.

Since $\cQ(2k,yA)\to1$ as $y\to0^+$ and tends to zero as $y\to\infty$,
the same integrability assumptions give
\begin{equation}
\mathcal F_k(0^+)=0,\qquad
\mathcal F_k(\infty)=1.
\label{eq:appendpoints}
\end{equation}
Integrating the derivative therefore gives the equivalent expression
\begin{equation}
\mathcal F_k(y)
=\frac{g^2}{\Gamma(2k)}
\int_0^y dw\,w^{2k-1}|\mathcal H_k(w)|^2.
\label{eq:appfrontintegrated}
\end{equation}
Thus the moving-front profile is the accumulated squared filtered
amplitude. Its derivative may diverge at $y=0$ while remaining integrable,
so the profile itself remains finite.

Finally, introduce the comoving coordinate
\begin{equation}
\tau=\alpha t-\frac12\log(4N),
\qquad y=e^{-2\tau}.
\end{equation}
The chain rule and Eq.~\eqref{eq:appfrontderivative} give
\begin{equation}
\partial_\tau\mathcal F_k(e^{-2\tau})
=-2y\mathcal F_k'(y)\leq0.
\label{eq:appmonotone}
\end{equation}
The scaling-limit profile therefore decreases monotonically with the
comoving coordinate $\tau$. At finite cut depth, however, the discrete
current can reverse sign, and the cumulative occupation can still exhibit
coherent rises and falls.


\subsection{Weighted-transform estimates at the profile edges}
\label{app:edgeestimates}

The two edges of the moving-front profile are controlled by different
parts of the boundary history $\xi_b$. The early edge, $y\to\infty$,
depends on emission near $s=0$, whereas the late edge, $y\to0^+$,
depends on the long-time decay of $\xi_b$. We estimate the filtered
amplitude $\mathcal H_k(y)$ in these regimes and then use
Eq.~\eqref{eq:appfrontintegrated} to obtain the corresponding profile
asymptotics.

\paragraph{Early edge.}
Since $|\xi_b(s)|\leq1$, the contribution from any interval
$s\geq s_0>0$ is exponentially suppressed relative to the endpoint
$s=0$ as $y\to\infty$. On the remaining interval, set $u=\alpha ys$
and use the early-time expansion in Eq.~\eqref{eq:boundaryearly}.
Endpoint expansion and dominated convergence then give
\begin{equation}
\mathcal H_k(y)
=\frac{(-i)^{d-1}U}{\alpha^d}
y^{-d}e^{-y/2}\,[1+O(y^{-1})].
\label{eq:appearlyHerror}
\end{equation}
Substituting this expression into Eq.~\eqref{eq:appfrontderivative} and
integrating from $y$ to infinity yields Eq.~\eqref{eq:frontleadingedge},
with relative error $O(y^{-1})$. All parameters, including $d$, are
held fixed in this limit.

\paragraph{Slow and oscillatory histories.}
Consider first a nonoscillatory history with
\begin{equation}
\xi_b(s)=A_be^{-\gamma s}+r(s),
\quad r(s)=o(e^{-\gamma s}),
\quad
p=k-\frac{\gamma}{2\alpha}>0.
\end{equation}
The leading exponential can be integrated exactly:
\begin{equation}
\int_0^\infty e^{(2k\alpha-\gamma)s}
e^{-ye^{2\alpha s}/2}\,ds
=\frac{1}{2\alpha}\left(\frac2y\right)^p
\Gamma(p,y/2).
\label{eq:appslowtransformexact}
\end{equation}
As $y\to0^+$, this is asymptotic to
$\Gamma(p)(2/y)^p/(2\alpha)$. To control the remainder, fix a large
$S$ and split the integral at $S$. The finite interval contributes
$O(1)=o(y^{-p})$, while for $s>S$ the bound
\[
|r(s)|\leq\varepsilon_S e^{-\gamma s},
\qquad \varepsilon_S\to0,
\]
bounds its contribution by $\varepsilon_S$ times the positive integral
in Eq.~\eqref{eq:appslowtransformexact}. Taking $y\to0^+$ and then
$S\to\infty$ establishes the leading coefficient in
Eq.~\eqref{eq:frontHpower}.

For a damped oscillatory pair, the same estimate applies to each
exponential component, with
\[
p_\pm=k-\frac{\delta_\gamma\pm i\nu_*}{2\alpha}.
\]
Combining the components before taking the squared modulus gives
Eq.~\eqref{eq:pairfront}, including its stated
$O(y^{\delta_\gamma})$ remainder. More generally, polynomial factors
$s^m$ in the history generate powers of $\log(1/y)$, as in
Eq.~\eqref{eq:frontmultiple}.

\paragraph{Marginal histories.}
At the marginal nonoscillatory threshold, suppose
\begin{equation}
\xi_b(s)=A_ms^m e^{-2k\alpha s}
+o(s^m e^{-2k\alpha s}),
\qquad m\geq0,\quad A_m\neq0.
\end{equation}
Set $L_y=\log(1/y)/(2\alpha)$. The filter in
Eq.~\eqref{eq:appfronttransform} is close to unity for $s\ll L_y$
and suppresses the integral for $s\gg L_y$. Splitting the integral into
the regions below, near, and above $L_y$ shows that the leading
contribution is $\int_0^{L_y}A_ms^m\,ds$. Hence
\begin{equation}
\mathcal H_k(y)\sim
\frac{A_m}{m+1}
\left[\frac{\log(1/y)}{2\alpha}\right]^{m+1}.
\label{eq:appmarginalH}
\end{equation}
Using Eq.~\eqref{eq:appfrontintegrated}, we obtain
\begin{equation}
\mathcal F_k(y)\sim
\frac{g^2|A_m|^2}
{2k\Gamma(2k)(m+1)^2(2\alpha)^{2m+2}}
y^{2k}[\log(1/y)]^{2m+2}.
\label{eq:appmarginalgeneral}
\end{equation}
For $m=0$, this reduces to Eq.~\eqref{eq:frontmarginal}.

\paragraph{Rapidly decaying histories.}
If $e^{2k\alpha s}\xi_b(s)$ is absolutely integrable, dominated
convergence gives
\begin{equation}
\mathcal H_k(y)\longrightarrow
H_0\equiv\int_0^\infty e^{2k\alpha s}\xi_b(s)\,ds.
\end{equation}
When $H_0\neq0$, Eq.~\eqref{eq:appfrontintegrated} yields
Eq.~\eqref{eq:frontfast}. If instead $H_0=0$, a sharper estimate
requires an additional moment condition. In particular, suppose
\begin{equation}
H_1=\int_0^\infty e^{2(k+1)\alpha s}\xi_b(s)\,ds
\end{equation}
converges absolutely. Expanding the filter to first order, with the
remainder controlled by this weighted integrability, gives
\begin{equation}
\mathcal H_k(y)=-\frac{yH_1}{2}+o(y).
\end{equation}
If $H_1\neq0$, it follows that
\begin{equation}
\mathcal F_k(y)\sim
\frac{g^2|H_1|^2}
{4(2k+2)\Gamma(2k)}y^{2k+2}.
\label{eq:appfastcancelled}
\end{equation}
A sufficient condition for absolute convergence of $H_1$ is
$|\xi_b(s)|\leq Ce^{-\gamma s}$ with
$\gamma>2(k+1)\alpha$. This stronger moment condition does not follow
from $H_0=0$ alone.

\paragraph{Constraint from the serial chain.}
For the serial chain considered here, the bare-tail poles
$z_j=-2i\alpha(k+j)$, $j=0,1,\ldots$, are zeros of
$[G_{\mathrm{UV}}(z)]_{b0}$ because
$\Delta_{d-1}(z_j)\neq0$. Whenever the corresponding integral converges
absolutely, this gives
\begin{equation}
H_j\equiv\int_0^\infty e^{2(k+j)\alpha s}\xi_b(s)\,ds
=i[G_{\mathrm{UV}}(z_j)]_{b0}=0.
\end{equation}
Thus the nonzero marginal histories considered above are excluded for
this chain. The rapidly decaying estimates
\eqref{eq:frontfast} and \eqref{eq:appfastcancelled} are therefore
conditional results, not necessarily the realized leading asymptotics.
Absolute convergence of $H_0$ implies
$\mathcal F_k(y)=O(y^{2k})$, while absolute convergence of $H_1$
implies $\mathcal F_k(y)=O(y^{2k+2})$ when $H_0=0$.
Identifying the first nonvanishing term requires further control of the
boundary history.


\section{Thermal response and weak-coupling matching}\label{app:bulk}

This appendix derives the Breitenlohner--Freedman (BF) scalar response,
constructs the measure of an independent oscillator coupled to that bath,
and proves the weak-coupling two-site and tail limits. The oscillator
coordinate $q$ is distinct from the scalar boundary operator
$\mathcal O$. The calculation assumes a fixed background, the stated
boundary-source convention, an all-frequency conformal bath, and either
a Gaussian bath or its leading-large-$N$ Gaussian approximation. No
backreaction, Schwarzian dynamics, or ultraviolet completion of a larger
geometry is inferred.

\subsection{Ingoing scalar solution and the BF response}
\label{app:bulkBF}

We briefly derive the thermal AdS$_2$ response at the BF bound. The general finite-$\nu$ result is standard~\cite{Faulkner}; here we focus on the $\nu\to0$ limit relevant for $k=\tfrac12$.

Consider a minimally coupled scalar 
on the fixed thermal AdS$_2$ geometry
\begin{equation}
d  s^2=-f(r)d  t^2+\frac{d  r^2}{f(r)},
\quad
f(r)=\frac{r^2-r_h^2}{\ell^2},
\quad
T=\frac{r_h}{2\pi\ell^2},
\label{eq:metric}
\end{equation}
with action
\begin{equation}
S_\Phi=-\frac{\mathcal A}{2}\int d  t\,d  r\sqrt{-g}
\bigl[(\nabla\Phi)^2+m^2\Phi^2\bigr],
\qquad \mathcal A>0.
\label{eq:probeaction}
\end{equation}
We use the source convention $H_{\rm src}=-J\mathcal O$ and
\begin{equation}
\chi_R(t)=- i \theta(t)\langle[\mathcal O(t),\mathcal O(0)]\rangle_\beta,
\quad
\chi_R(z)=\int e^{ i  zt}\chi_R(t)\,d  t,
\label{eq:appretarded}
\end{equation}
so that the retarded response is the negative of the source-to-expectation-value kernel.

For $\Phi=e^{- izt}\phi_z(r)$, setting
\begin{equation}
 x=\frac r{r_h},
\qquad
\alpha=\pi T,
\qquad
k=\frac12+\nu,
\qquad
\mu=\frac{ i  z}{2\alpha},   
\end{equation}

the radial equation becomes
\begin{equation}
(x^2-1)\phi_z''+2x\phi_z'
+\left[\frac{(z/2\alpha)^2}{x^2-1}-k(k-1)\right]\phi_z=0.
\label{eq:appradialequation}
\end{equation}
The ingoing solution is~\cite{SonStarinets,KovtunStarinets,Faulkner}
\begin{equation}
\phi_z(x)=\frac{\left(\frac{x+1}{x-1}\right)^{\frac\mu2}}
{\Gamma(1-\mu)}
{}_2F_1\left(1-k,k;1-\mu;\frac{1-x}{2}\right).
\label{eq:appradialsolution}
\end{equation}
Near the boundary,
\begin{equation}
\phi_z(r)=A_\nu(z)r^{-1/2+\nu}
+B_\nu(z)r^{-1/2-\nu}+\cdots ,
\label{eq:appfalloffs}
\end{equation}
with
\begin{equation}
\frac{B_\nu(z)}{A_\nu(z)}
=(2r_h)^{2\nu}
\frac{\Gamma(-2\nu)\Gamma(1/2+\nu)}
{\Gamma(2\nu)\Gamma(1/2-\nu)}
\frac{\Gamma(1/2+\nu- iz/(2\alpha))}
{\Gamma(1/2-\nu- iz/(2\alpha))}.
\label{eq:appradialratio}
\end{equation}

At the BF bound, $\nu\to0$, the two falloffs degenerate and combine into a logarithmic solution. Introducing dimensionless coefficients
\begin{equation}
\bar A_\nu=\ell^\nu A_\nu,
\qquad
\bar B_\nu=\ell^{-\nu}B_\nu,    
\end{equation}

one finds
\begin{equation}
\phi_z(r)=r^{-1/2}
\left[A_{\BF}(z)\log(r/\ell)+B_{\BF}(z)+\cdots\right],
\label{eq:appBFfalloffs}
\end{equation}
where, for example,
\begin{equation}
A_{\BF}(z)
=\lim_{\nu\to0}\nu(\bar A_\nu-\bar B_\nu)
=\sqrt{\frac{2r_h}{\pi}}\,
\frac1{\Gamma(1/2- iz/(2\alpha))}.
\label{eq:appBFsourceexplicit}
\end{equation}

Expanding the finite-$\nu$ ratio to first order gives
\begin{align}
& \frac{\bar B_\nu(z)}{\bar A_\nu(z)}
=-[1+2\nu c(z)+O(\nu^2)],
\nonumber\\
&c(z)=\log\frac{2r_h}{\ell}
+2\gamma_E+\Psi(1/2)
+\Psi\!\left(\frac12-\frac{ i  z}{2\alpha}\right).
\label{eq:appBFcexpansion}   
\end{align}
The leading $-1$ simply reflects the degeneracy of the two falloffs. The finite response is encoded in the first-order term. Since
\begin{equation}
  \frac{B_{\BF}(z)}{A_{\BF}(z)}
 =-c(z),  
\end{equation}
a static subtraction removes all frequency-independent terms and gives
\begin{align}
&\frac{B_{\BF}(z)}{A_{\BF}(z)}
-\frac{B_{\BF}(0)}{A_{\BF}(0)}
=-h_\alpha(z),
\nonumber\\
&h_\alpha(z)
=\Psi\left(\frac12-\frac{ i  z}{2\alpha}
\right)-\Psi\left(\frac12\right).
\label{eq:appBFfiniteratio}    
\end{align}
The reflection formula implies
\begin{equation}
h_\alpha(0)=0,
\qquad
\Im h_\alpha(\omega)
=-\frac{\pi}{2}
\tanh\left(\frac{\pi\omega}{2\alpha}\right).
\label{eq:appImh}
\end{equation}

At the BF point the logarithmic falloff requires holographic renormalization. Using the standard logarithmic counterterm and varying the renormalized on-shell action gives, up to a real local source term,
\begin{equation}
\langle\mathcal O\rangle_J
=\frac{\mathcal A}{\ell^2}B_{\BF}.
\label{eq:appBFvev}
\end{equation}
With the retarded-response convention above, the statically subtracted response is therefore
\begin{equation}
\chi^R_{\BF}(z)-\chi^R_{\BF}(0)
=\mathcal C_{\BF}h_\alpha(z),
\qquad
\mathcal C_{\BF}=\frac{\mathcal A}{\ell^2}>0.
\label{eq:appBFactionresponse}
\end{equation}
Thus the thermal AdS$_2$ scalar at the BF bound produces the digamma response whose imaginary part is proportional to
\begin{equation}
 -\tanh\!\left(\frac{\pi\omega}{2\alpha}\right).   
\end{equation}
This is the $k=\tfrac12$ member of the conformal family used in the main text.


\subsection{Retarded elimination of the oscillator bath}
\label{app:oscillator}

The independent oscillator and its coupling to the bath are described by
\begin{equation}
S_q=\frac12\int  dt\,\bigl(\dot q^2-\Omega_0^2q^2\bigr),
\qquad
S_{\rm int}=\lambda_0\int  dt\,q\mathcal O.
\label{eq:apposcaction}
\end{equation}
A force coupled through $H_{\rm src}=-Jq$ gives
\begin{equation}
 \ddot q+\Omega_0^2q=J+\lambda_0\langle\mathcal O\rangle.   
\end{equation}
The bath is sourced by $\lambda_0q$, so its induced expectation value is
\begin{equation}
 \langle\mathcal O\rangle(t)
=-\lambda_0\int  dt'\,\chi^R(t-t')q(t').   
\end{equation}
Eliminating the bath therefore gives
\begin{equation}
\bigl[-z^2+\Omega_0^2+\lambda_0^2\chi^R(z)\bigr]q(z)=J(z).
\label{eq:appbareosceom}
\end{equation}
For a Gaussian bath, this retarded self-energy is exact; at leading large $N$, the same result follows when higher connected bath correlators are suppressed. The corresponding causal influence functional is standard. We use the doubled-field formulation of Ref.~\cite{AppendicesKamenev} and do not reproduce it here.

Absorb the positive bath coefficient and its real static contribution into
\begin{equation}
\lambda^2=\lambda_0^2\mathcal C_{\BF},
\qquad
\Omega_r^2=\Omega_0^2+\lambda_0^2\chi^R_{\BF}(0),
\label{eq:appeffaction2}
\end{equation}
so that
$
\lambda_0^2\chi^R_{\BF}(z)
=\Omega_r^2-\Omega_0^2+\lambda^2h_\alpha(z).
$

The equation of motion then becomes
\begin{equation}
\bigl[-z^2+\Omega_r^2+\lambda^2h_\alpha(z)\bigr]q(z)=J(z),
\label{eq:apposceom}
\end{equation}
with solution
\begin{equation}
q(z)=-D_R(z)J(z),
\qquad
D_R(z)=\frac1{z^2-\Omega_r^2-\lambda^2h_\alpha(z)}.
\label{eq:appDR}
\end{equation}
We assume $\Omega_r>0$. In this normalization, both $\lambda$ and $\Omega_r$ have dimensions of frequency. The sign in $q=-D_RJ$ gives the correct static displacement, $q=J/\Omega_r^2$.


\subsection{Weak-coupling limit of the oscillator measure}
\label{app:matching}

Throughout this subsection, $\Omega_r>0$ and $\alpha>0$ are fixed. We
take $\lambda\to0^+$ with $\epsilon=\lambda^2$, keeping every polynomial
degree and Krylov depth fixed before taking the limit. Introduce
\begin{equation}
X(\omega)=\omega^2-\Omega_r^2,
\qquad
S(\omega)=\sech\left(\frac{\pi\omega}{2\alpha}\right),
\label{eq:appweaknotation}
\end{equation}
and write
\begin{equation}
h_\alpha(\omega)=u(\omega)+iv(\omega),
\end{equation}
where $u$ is even, $v$ is odd, and $v(\Omega_r)<0$. The unnormalized
probe measure is
\begin{equation}
\widetilde\rho_q(\omega)
=
\frac{\epsilon S(\omega)}
{4\bigl[X(\omega)-\epsilon u(\omega)\bigr]^2
+4\epsilon^2v(\omega)^2}.
\label{eq:appproberho}
\end{equation}
We first determine its weak-coupling limit and the leading behavior of
the first two Krylov bonds. We then use a quadratic spectral filter to
identify the measure seen from site $2$ and recover the conformal tail.
Finally, we compute the leading correction to the first bond from a
signed spectral moment.


\subsubsection{The weak-coupling limit and the first two Krylov bonds}

As $\lambda\to0^+$, the spectral weight concentrates near
$\omega=\pm\Omega_r$. The resonances have widths of order
$\epsilon/\Omega_r$, so the detuning must be retained in the denominator
when evaluating their integrated weights. Near the positive-frequency
resonance, set
\begin{equation}
x=\omega-\Omega_r,
\qquad
d_r=-v(\Omega_r)>0.
\end{equation}
Since
\begin{equation}
X(\omega)=2\Omega_r x+O(x^2),
\end{equation}
the density has the local Lorentzian form
\begin{align}
\widetilde\rho_q(\Omega_r+x)
&\simeq
\frac{\epsilon S(\Omega_r)}{16\Omega_r^2}
\frac{1}{(x-\Delta\Omega)^2+\Gamma_\lambda^2},
\nonumber\\
\Delta\Omega
&=\frac{\epsilon u(\Omega_r)}{2\Omega_r},
\qquad
\Gamma_\lambda
=-\frac{\epsilon v(\Omega_r)}{2\Omega_r}
=\frac{\epsilon d_r}{2\Omega_r}.
\label{eq:applorentzian}
\end{align}
Both the resonance shift and the width are of order $\epsilon$ and must
therefore be retained in the resonance integral.

Using
\begin{equation}
v(\Omega_r)
=-\frac{\pi}{2}\tanh\left(\frac{\pi\Omega_r}{2\alpha}\right),
\qquad
S(\Omega_r)
=\sech\left(\frac{\pi\Omega_r}{2\alpha}\right),
\end{equation}
we find that the positive-frequency resonance carries weight
\begin{align}
\int_{-\infty}^{\infty}
\widetilde\rho_q(\Omega_r+x)\,dx
&\simeq
\frac{\epsilon S(\Omega_r)}{16\Omega_r^2}
\frac{\pi}{\Gamma_\lambda}
\nonumber\\
&=
\frac{1}{4\Omega_r
\sinh[\pi\Omega_r/(2\alpha)]}.
\label{eq:appatomweight}
\end{align}
The negative-frequency resonance has the same weight. Away from these
resonances, the density vanishes pointwise as $\epsilon\to0$, while
the factor $S(\omega)$ suppresses the high-frequency tails. Separating
the two resonant neighborhoods from their complement therefore gives
the weak limit
\begin{equation}
\widetilde\rho_q(\omega)\,d\omega
\ \rightharpoonup\
\frac{\delta(\omega-\Omega_r)+\delta(\omega+\Omega_r)}
{4\Omega_r\sinh[\pi\Omega_r/(2\alpha)]}\,d\omega .
\label{eq:appfreelimit}
\end{equation}
Consequently,
\begin{align}
N_q&\longrightarrow
N_0=
\frac{1}{2\Omega_r\sinh[\pi\Omega_r/(2\alpha)]},
\nonumber\\
\rho_q(\omega)\,d\omega
&\ \rightharpoonup\
\frac12\left[
\delta(\omega-\Omega_r)+\delta(\omega+\Omega_r)
\right]\,d\omega .
\label{eq:appN0}
\end{align}
Here $\rho_q=\widetilde\rho_q/N_q$ is the normalized measure. Its
limiting two-atom support implies
\begin{equation}
b_1^q\longrightarrow\Omega_r,
\qquad
b_2^q\longrightarrow0.
\end{equation}
This is the strictly limiting measure; at every nonzero coupling, the
chain still contains the conformal tail.

To extract the tail, we remove the resonant contribution using the
quadratic filter $X(\omega)^2$. Equation~\eqref{eq:appproberho} gives
\begin{equation}
\frac{X(\omega)^2}{\epsilon}\widetilde\rho_q(\omega)
=
\frac{S(\omega)}4
\frac{X(\omega)^2}
{[X(\omega)-\epsilon u(\omega)]^2+\epsilon^2v(\omega)^2}.
\label{eq:appfiltered}
\end{equation}
For fixed $\omega\neq\pm\Omega_r$, the ratio on the right tends to
one. The contribution from the shrinking resonant neighborhoods
vanishes after multiplication by $X^2/\epsilon$, and the exponential
decay of $S$ controls the high-frequency region. Thus, for every fixed
polynomial $p$,
\begin{equation}
\lim_{\lambda\to0^+}
\int p(\omega)\frac{X(\omega)^2}{\lambda^2}
\widetilde\rho_q(\omega)\,d\omega
=
\frac14\int p(\omega)S(\omega)\,d\omega .
\label{eq:appfilteredmoments}
\end{equation}
Equivalently, the filtered measures converge in every fixed polynomial
moment:
\begin{equation}
\frac{X(\omega)^2}{\lambda^2}
\widetilde\rho_q(\omega)\,d\omega
\ \longrightarrow\
\frac14 S(\omega)\,d\omega
=
\frac{\alpha}{2}\rho_{1/2}(\omega)\,d\omega .
\label{eq:appfilteredlimit}
\end{equation}

We next determine the leading interface bond. Define the centering
correction and a logarithmic scale by
\begin{equation}
\delta\mu=\mu_2-\Omega_r^2,
\qquad
L_\epsilon=
1+\left|\log\left(\frac{\epsilon}{\alpha^2}\right)\right|.
\end{equation}
The resonance neighborhoods and the intermediate region
$\epsilon\ll|X|\ll1$ give the bound
\begin{equation}
\delta\mu=O(\epsilon L_\epsilon),
\qquad
(\delta\mu)^2
=O(\epsilon^2L_\epsilon^2)=o(\epsilon).
\label{eq:appcenteringbound}
\end{equation}
The high-frequency contribution is exponentially suppressed by
$S(\omega)$.

For the normalized even measure, let
\begin{equation}
\mu_2=\int\omega^2\rho_q(\omega)\,d\omega,
\qquad
\mu_4=\int\omega^4\rho_q(\omega)\,d\omega.
\end{equation}
Expanding $\omega^2=X+\Omega_r^2$ gives
\begin{equation}
\mu_4
=\Omega_r^4
+2\Omega_r^2(\mu_2-\Omega_r^2)
+\int X^2\rho_q(\omega)\,d\omega .
\label{eq:appmu4}
\end{equation}
Using the filtered limit and $N_q\to N_0$, we obtain
\begin{align}
\int X^2\rho_q(\omega)\,d\omega
&=
\lambda^2\frac{\alpha}{2N_0}
+o(\lambda^2)
\nonumber\\
&=
\lambda^2\alpha\Omega_r
\sinh\left(\frac{\pi\Omega_r}{2\alpha}\right)
+o(\lambda^2).
\label{eq:appmu4filtered}
\end{align}
For a symmetric normalized measure, the first two Lanczos coefficients
satisfy
\begin{equation}
(b_1^q)^2=\mu_2,
\qquad
(b_1^q)^2(b_2^q)^2=\mu_4-\mu_2^2.
\end{equation}
It follows from Eq.~\eqref{eq:appmu4} that
\begin{align}
\mu_4-\mu_2^2
&=
\int X^2\rho_q(\omega)\,d\omega-(\delta\mu)^2
\nonumber\\
&=
\lambda^2\frac{\alpha}{2N_0}
+o(\lambda^2).
\label{eq:appb1b2}
\end{align}
Hence
\begin{equation}
b_1^q\longrightarrow\Omega_r,
\qquad
\frac{(b_2^q)^2}{\lambda^2}
\longrightarrow
B_\alpha(\Omega_r)
\equiv
\frac{\alpha}{\Omega_r}
\sinh\left(\frac{\pi\Omega_r}{2\alpha}\right).
\label{eq:appBalpha}
\end{equation}
Since $b_2^q>0$,
\begin{equation}
b_2^q
=\lambda\sqrt{B_\alpha(\Omega_r)}+o(\lambda).
\end{equation}
Thus the leading interface strength follows from fixed polynomial
moments alone; no higher-order expansion of the narrowing resonance
integrals is needed.


\subsubsection{From the site measure to the shifted tail}
\label{app:Ptwo}

Let $\mathcal L=[H,\cdot]$ and
$|K_0)=|q)/\sqrt{N_q}$. The next two Krylov vectors are
\begin{equation}
|K_1)
=\frac{|[H,q])}{b_1^q\sqrt{N_q}},
\qquad
|K_2)
=\frac{(\mathcal L^2-\mu_2)|K_0)}
{b_1^qb_2^q},
\label{eq:appqKrylovvectors}
\end{equation}
where the subtraction removes the component along $|K_0)$. Since
$\rho_q$ is even, $|K_2)$ is also orthogonal to $|K_1)$. The associated
degree-two orthogonal polynomial and site-$2$ spectral measure are
\begin{equation}
P_2^q(\omega)
=\frac{\omega^2-\mu_2}{b_1^qb_2^q},
\qquad
d\rho_q^{(2)}(\omega)
=\bigl(P_2^q(\omega)\bigr)^2\rho_q(\omega)\,d\omega.
\label{eq:appPtwo}
\end{equation}

We may replace $\omega^2-\mu_2=X-\delta\mu$ by $X$ in the filtered
measure at leading order. Indeed, for every fixed polynomial $p$, the
moment bounds
\begin{align}
&\int |p(\omega)X(\omega)|\rho_q(\omega)\,d\omega
=O(\epsilon L_\epsilon),
\nonumber\\
&\int |p(\omega)|\rho_q(\omega)\,d\omega=O(1),
\end{align}
and $\delta\mu=O(\epsilon L_\epsilon)$ imply
\begin{align}
&\frac1\epsilon
\left|
\int p(\omega)
\bigl[(X-\delta\mu)^2-X^2\bigr]\rho_q(\omega)\,d\omega
\right|
\nonumber\\
&\quad\leq
\frac{2|\delta\mu|}{\epsilon}
\int |pX|\rho_q\,d\omega
+
\frac{(\delta\mu)^2}{\epsilon}
\int |p|\rho_q\,d\omega
\nonumber\\
&\quad=O(\epsilon L_\epsilon^2)\longrightarrow0.
\label{eq:appcenteredfilterproof}
\end{align}
Thus the centering correction does not affect the limiting filtered
polynomial moments.

Combining Eqs.~\eqref{eq:appfilteredlimit},
\eqref{eq:appb1b2}, and $N_q\to N_0$, we conclude that
\begin{equation}
d\rho_q^{(2)}(\omega)
\ \longrightarrow\
\rho_{1/2}(\omega)\,d\omega
\label{eq:appsite2}
\end{equation}
in every fixed polynomial moment. The spectral measure seen from
global site $2$ therefore converges to the BF measure.

This limiting measure is distinct from the two-atom measure seen from
sites $0$ and $1$. Its moments are those of the Krylov chain beginning
at global site $2$:
\begin{equation}
v_{2r}
=\int\omega^{2r}\,d\rho_q^{(2)}(\omega)
=\langle e_2|K_q^{2r}|e_2\rangle.
\label{eq:appsitewalkmoments}
\end{equation}
In particular,
\begin{equation}
\begin{aligned}
v_2
&=(b_2^q)^2+(b_3^q)^2,
\\
v_4
&=(b_1^qb_2^q)^2
+\bigl[(b_2^q)^2+(b_3^q)^2\bigr]^2
+(b_3^qb_4^q)^2.
\end{aligned}
\label{eq:appfirstwalkmoments}
\end{equation}
The BF limits $v_2\to\alpha^2$ and $v_4\to5\alpha^4$, together with
$b_2^q\to0$, imply successively that
$b_3^q\to\alpha$ and $b_4^q\to2\alpha$.

More generally, for each fixed $r$, the unique closed walk of length
$2r$ from site $2$ that reaches the furthest site $r+2$ contributes
the product of bonds along the outward and return paths. Thus
\begin{equation}
v_{2r}
=
\left(\prod_{j=3}^{r+2}b_j^q\right)^2
+V_r(b_1^q,\ldots,b_{r+1}^q),
\label{eq:appwalkinduction}
\end{equation}
where $V_r$ collects the remaining walks. Any contribution that reaches
site $1$ or $0$ contains the factor $(b_2^q)^2$ and vanishes in the
weak-coupling limit. The remaining terms involve only previously
determined bonds. The BF moments therefore determine the limits
successively:
\begin{equation}
b_{n+2}^q\longrightarrow b_n^{\mathrm T}
=\alpha n,
\qquad n\geq1\ \text{fixed}.
\label{eq:apptail}
\end{equation}
Hence global sites $0$ and $1$ form the shallow sector, global site $2$
becomes tail site $0_T$, and $b_2^q$ is the interface bond. The
two-site shift is a relabeling of the chain beyond the shallow sector.


\subsubsection{The signed first moment and the first-bond shift}
\label{app:firstbondproof}

The quadratic filter fixes the leading interface strength, but it does
not determine the order-$\lambda^2$ correction to the internal bond
$b_1^q$. This correction is encoded in the signed moment
\begin{equation}
\frac1{\lambda^2}
\int X(\omega)\rho_q(\omega)\,d\omega,
\qquad
X(\omega)=\omega^2-\Omega_r^2.
\label{eq:appsignedmoment}
\end{equation}
Unlike the squared filter, this moment retains the sign of $X$ and
therefore distinguishes the dispersive shift of the resonance from
the effect of its width.

A termwise expansion of
\[
[X(\omega)-\epsilon u(\omega)]^2+\epsilon^2v(\omega)^2
\]
in powers of $\epsilon$ is not uniform near $X=0$, where the
denominator becomes small. Instead, we treat the two resonances using
the distributional limit
\begin{equation}
\frac{X}
{(X-\epsilon u)^2+\epsilon^2v^2}
\ \longrightarrow\
\operatorname{PV}\frac1X
+\pi\frac{u}{|v|}\delta(X).
\label{eq:appsigneddistribution}
\end{equation}
Here the identity is understood distributionally, with $u$ and $v$
evaluated at the zeros of $X$. To see its origin, write
$X=(X-\epsilon u)+\epsilon u$. The first term yields the principal-value
distribution, while the second produces a delta sequence centered at
$X=\epsilon u$. The Jacobian associated with changing variables from
$\omega$ to $X$ is included when the distributions act on a test
function.

Multiplying Eq.~\eqref{eq:appsigneddistribution} by $S(\omega)/4$ and
integrating over both resonances gives
\begin{align}
\lim_{\lambda\to0^+}
\frac1{\lambda^2}
\int X\widetilde\rho_q(\omega)\,d\omega
&=
\frac{\alpha}{2}
\operatorname{PV}\int_{\mathbb R}
\frac{\rho_{1/2}(\omega)}
{\omega^2-\Omega_r^2}\,d\omega
\nonumber\\
&\quad+N_0\Re h_\alpha(\Omega_r).
\label{eq:appunnormalizedsigned}
\end{align}
The delta contributions at $\omega=\pm\Omega_r$ combine to give
$N_0u(\Omega_r)$, while the principal-value term describes the
dispersive response of the continuum. Both contributions are needed
to determine the signed moment.

Using the evenness of $\rho_{1/2}$, we have
\begin{equation}
\operatorname{PV}\int_{\mathbb R}
\frac{\rho_{1/2}(\omega)}
{\omega^2-\Omega_r^2}\,d\omega
=
-\frac{\Re m_{1/2}^+(\Omega_r)}{\Omega_r},
\label{eq:appPVidentity}
\end{equation}
where the identity follows by averaging
$1/(\Omega_r-\omega)$ and $1/(\Omega_r+\omega)$ in the real part of
the Jacobi transform. Dividing Eq.~\eqref{eq:appunnormalizedsigned} by
$N_q\to N_0$ then yields
\begin{align}
\mu_2-\Omega_r^2
={}&
\lambda^2
\bigl[
\Re h_\alpha(\Omega_r)
-\Omega_r B_\alpha(\Omega_r)
\Re m_{1/2}^+(\Omega_r)
\bigr]\nonumber\\
+&o(\lambda^2).
\label{eq:appsignedlimit}
\end{align}
This improves the earlier bound
$O(\lambda^2L_{\lambda^2})$ and identifies the leading coefficient.
Taking the square root of $(b_1^q)^2=\mu_2$ gives
\begin{align}
b_1^q
&=\Omega_r+\lambda^2\Delta b_1+o(\lambda^2),
\nonumber\\
\Delta b_1
&=
\frac{
\Re h_\alpha(\Omega_r)
-\Omega_r B_\alpha(\Omega_r)
\Re m_{1/2}^+(\Omega_r)}
{2\Omega_r}.
\label{eq:appdeltau}
\end{align}

\bibliographystyle{apsrev4-2}
\bibliography{literature}

\end{document}